\documentclass[a4paper,11pt]{article}
\usepackage{times}
\usepackage{amsmath,amssymb, graphicx,epsfig,here}
\usepackage{gensymb}
\usepackage{subcaption}
\usepackage{xfrac}
 \usepackage[colorlinks=true]{hyperref}
\hypersetup{urlcolor=blue, citecolor=red}
\providecommand{\keywords}[1]
{
  \small	
  \textbf{\textit{Keywords---}} #1
}

\begin{document}
\title{Modelling with Volna-OP2: Towards tsunami threat reduction}
\author{Daniel Giles$^{1}$, Brian McConnell $^{3}$ and Fr\'ed\'eric Dias $^{1,2}$\\
{\footnotesize $^{1}$ School of Mathematics and Statistics, University College Dublin, Dublin, Ireland} \\
{\footnotesize $^{2}$ Earth Institute, University College Dublin, Dublin, Ireland}\\
{\footnotesize $^{3}$ Geological Survey Ireland, Dublin, Ireland}}



\maketitle
\abstract{Accurate and efficient tsunami modelling is essential for providing tsunami forecasts and hazard assessments. Volna-OP2 is a finite volume solver of the nonlinear shallow water equations and its capabilities of producing both faster than real time ensembles and high resolution inundation studies are presented here. The code is massively parallelised and can utilise various high performance computing architectures. When an earthquake is detected there is always some uncertainty on the source parameters. Generating a faster than real time ensemble for maximum wave heights which captures this uncertainty would be of great benefit to tsunami warning centres. The 2003 Boumerdes earthquake (Algeria) acts as a test case for showing Volna-OP2's ability at rapidly forecasting regional maximum wave heights. Drawing on various earthquake sources proposed in the literature and scaling the magnitudes to mimic uncertainty on the source, 20 separate earthquake realisations are simulated for 4 hours real time in 97s on two Nvidia V100 GPUs. Further a reduced ensemble of the Lisbon 1755 tsunami with an emphasis on the effects to the Irish coastline is presented. Where again various earthquake sources have been drawn from the literature and simulated on a regional scale. Finally, a pilot study which builds upon the reduced ensemble results investigates the inundation of a Lisbon tsunami on key sections of the Irish coastline. The results of this pilot study highlight that the inundation is constrained to low-lying areas with a maximum run-up heights of $\approx 3.4m$ being found.\\}



\section{Introduction}
Tsunamis are infrequent events that have the capacity of being extremely destructive. Owing to their rarity the tsunami community relies heavily on mathematical and computational modelling to provide accurate forecasts and risk assessments. Tsunamis can travel at extremely high speeds in deep water, thus arrival times at the coastline are often on the $\mathcal{O}$(mins). Tsunami Warning Centres (TWC) are therefore under severe time constraints when deciding the level of warning and areas at risk from an event. To aid in this decision process, TWCs currently deploy a variety of tools which produce regional levels of warnings. 

However, it is well documented that the local bathymetry can have a substantial effect on the local levels of maximum wave heights, inundation and induced currents. To gain an understanding of how the local bathymetry affects an incoming tsunami, prior high resolution risk assessments are often carried out. These studies require extensive computational resources and are time consuming. Epistemic uncertainty involved in the modelling and source parameters has lead to the development of probabilistic tsunami hazard assessments (PTHA)\cite{Geist2006, Grezio2017}. One of the key limiting factors for both TWCs and risk assessment studies is the inherent tsunami modelling software. Extremely efficient tsunami codes, which utlitize modern computational techniques, have been developed to reduce the run times involved. 

The software package Volna-OP2 is a tsunami code which has been optimised to run on various high performance architectures, CPUs, Xeon-Phis and GPUs. To date, Volna-OP2 has been used by various research groups around the world. In particular its  efficiency has been leveraged in conjunction with statistical studies to carry out computationally expensive tasks like sensitivity studies and building statistical surrogates \cite{Salmanidou2017, Guillas2018, Gopinathan2020}. The code solves the depth-averaged nonlinear shallow water equations, utilising modern reconstruction and shock capturing techniques. The original code was introduced in 2011 \cite{Dutykh2011} and has since undergone extensive parallelisation \cite{Reguly2018}. Both the original and newly updated version have been verified against the traditional tsunami mitigation benchmark problems and an extensive error analysis of the newly parallelised version has been carried out \cite{Giles2020}.

Volna-OP2 is presented as a candidate code which could be implemented into a TWC workflow. It is shown here to be capable of rapidly simulating tsunamis on a regional scale which can then be incorporated into 'on the fly' ensemble warnings. Further, to demonstrate the code's capability at performing high resolution inundation studies, a pilot mapping study of the Irish coastline has been carried out.  This paper is thus split into three parts; in section(\ref{Numerics}) an outline of the numerical scheme and parallelisations used in Volna-OP2 are described. In section(\ref{FRTE}) a faster than real time tsunami ensemble is presented for the Mediterranean Sea and a reduced ensemble for the North East Atlantic Ocean. In section (\ref{Inundation}) a high resolution inundation mapping pilot study of the Irish coastline is shown. The main tsunami threat for the Irish coastline lies with a Lisbon type earthquake or a mid-Atlantic submarine landslide. This pilot study focuses on the Lisbon 1755 tsunami as a reasonable worst case scenario and builds upon the reduced ensemble results. 

\section{Numerical scheme and parallelisation}\label{Numerics}

In the deep ocean tsunamis exhibit long wavelengths when compared to the depth. Thus, the simplest approach in modelling tsunami waves is to invoke the shallow water limit and to model the tsunami dynamics using the linear shallow water equations. These linear equations can be simply dealt with using computational techniques and have been shown to work well for modelling tsunamis across a deep ocean basin. However, once the ratio between the depth and wave height becomes comparable, nonlinear dynamics play an important role and the nonlinear shallow water equations are needed to represent the tsunami dynamics. Continuing on this trend, if one is to include dispersion or non-hydrostatic pressures one gets the Boussinesq equations. There are plenty of numerical solvers that exist which tackle these higher order Boussinesq equations. Undoubtedly, physical dispersion can play an important role but in most cases the nonlinear shallow water equations are sufficient. 

Volna-OP2 is a finite-volume solver which solves the depth-average nonlinear shallow water equations,

\begin{equation}
\begin{aligned}
    \frac{\partial H}{\partial t}+ \nabla\cdotp (H\vec{u}) &= 0,\\
    \frac{\partial(H\vec{u})}{\partial t}+ \nabla \cdotp (H\vec{u}\otimes\vec{u} +\frac{g}{2}H^2 \textbf{I})&= - gH\nabla h,\label{eqs}
\end{aligned}
\end{equation}
    where $ H=(b + \eta) $ is the total water depth,  $ b(x,y,t) $ is the time-dependent bathymetry, $\eta(x,y,t) $ is the free surface deformation, $ \vec{u}(u,v) $ is the depth averaged fluid velocity in the $ x $ and $ y $ horizontal directions, \textbf{I} is the identity matrix and $ g $ is the acceleration due to gravity.
    
Volna-OP2 utilises unstructured triangular meshes with dependent variables assigned to the triangle's barycenters. In order to achieve 2$^{nd}$ order accuracy in space, which is essential in modern methods, a MUSCL reconstruction scheme is implemented.  For the temporal discretisation an option of using either a 2$^{nd}$ or 3$^{rd}$  order strong stability preserving Runge-Kutta scheme is present. Extensive details on the numerical scheme used in Volna-OP2 can be found in previous works \cite{Dutykh2011, Giles2020}. 

When Volna was first developed in 2011 \cite{Dutykh2011} it was naturally a serial code. However, the code underwent a series of parallelisations culminating in the present Volna-OP2 version. The main parallelisation of the serial code occurred when it was built on the OP2 domain-specific language (DSL) for unstructured mesh computations \cite{Mudalige2012}. This DSL enables unstructured mesh calculations to be expressed at a high level, with a suite of automated tools to translate the scientific code into a wide range of targeted high performance implementations.  OP2 makes use of various parallelisation approaches: MPI, OpenMP and CUDA. It allows for the Volna numerical algorithms to be written once, which in turn is then automatically parallelised to use multiple CPUs and GPUs. This parallelisation of Volna resulted in a huge improvement in the codes scalability and performance \cite{Reguly2018, Giles2020}. It has enabled the code to be used for various extremely computationally expensive tasks like building statistical emulators \cite{Beck2016, Liu2017, Guillas2018, Beck2016, Gopinathan2020}, carrying out sensitivity analysis  and stochastic inversions \cite{Gopinathan2017}. Extensive details on the parallelisation approaches used in OP2 can be found here \cite{Giles2011, Mudalige2012} and details on the integration of Volna into OP2 are given in \cite{Reguly2018}.

\section{Faster than Real Time Ensemble}\label{FRTE}

As stated TWCs are tasked with deciding the level of threat posed by a tsunamigenic event and the areas at risk. Warning centres utilise various different tools to aid in this decision process. Leveraging modern high performance computing resources and improvements in the accuracy of earthquake detection has allowed one to simulate tsunami's across an ocean basin with run times which are faster than real time (FTRT) \cite{Lovholt2019}. The results of these faster than real time simulations are supplemental to the traditional decision matrix and precomputed database techniques that are currently utilised \cite{Gailler2013}.  However, when an earthquake is detected, there is initially uncertainty with respect to various physical parameters of the event. Thus, to capture this uncertainty, the main goal is to produce faster than real time ensemble techniques. With arrival times on the order of minutes it is imperative that these ensemble results are completed in a timely fashion. Workflows are currently being developed on the implementation of these highly efficient codes into an automatically triggered response to seismic activity \cite{Lovholt2019}. In the following section Volna-OP2's ability to provide extremely rapid regional forecasts and a brief example of a potential ensemble prediction system is shown with the Mediterranean as a test case. 

\subsection{Mediterranean Tsunami}
By way of example the major earthquake of Boumerdes 2003 (6.8 Mw) is chosen. The earthquake has been extensively studied in the literature \cite{Bounif2004, Meghraoui2004, Yelles2004, Semmane2005}. The exact source parameters from various papers vary and will be used to highlight the ability of Volna-OP2 to provide faster than real time ensemble results. As stated in the immediate aftermath of an earthquake there is always uncertainty on the source parameters and magnitude. To capture this uncertainty, the earthquake sources presented in the literature have been scaled using Wells and Coppersmith \cite{Wells1994} relations to provide initial earthquakes with magnitudes varying between 6.7- 7.1 Mw. In total, there are 20 different initial earthquake sources. The initial water surface elevation is obtained by using the Okada Model \cite{Okada1985}. 

Maximum wave heights at each grid point are outputted after 4 hours of simulated real time. Absolute maximum and mean wave heights which incorporate the complete set are then calculated and outputted. The resolution of the grid is 2arc minutes and is thus too coarse for direct comparison to tidal gauge measurements. This grid resolution results in $\approx$ 385,000 cells. The run time for the complete set of 20 separate 4 hour simulations is 97s, using 2 Nvidia V100 GPUs. It should be noted that this is the run time for just the simulations themselves and doesn't include the time taken to post-process the data. Further work on these workflows is required and has been previously highlighted \cite{Lovholt2019}.

\begin{figure*}[h!]
	\centering
	\includegraphics[width = 15cm, height = 10cm]{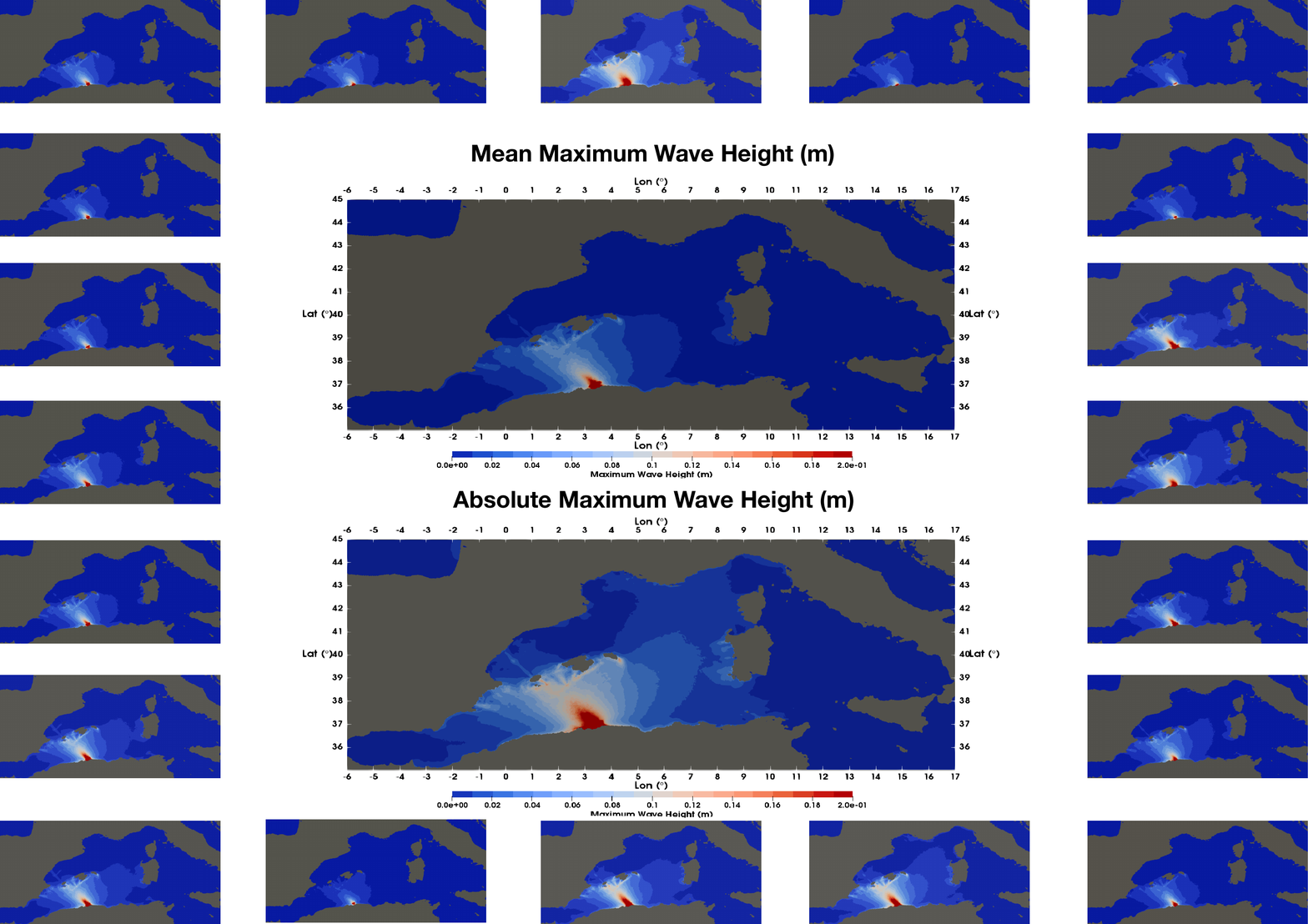}
	\caption[Med_warning]{Outside figures are the maximum wave heights from each of the 20 realisations. Centre figures are the combined absolute maximum wave heights and mean maximum wave heights.} 
	\label{Med_warning}
\end{figure*}

From the results shown it can be seen that the larger magnitude earthquakes produce larger maximum wave heights. The regional areas which are worst affected are consistent across all the realisations, ie. mainly the local Algerian and the Balearic coastlines. The mean maximum wave heights can be deemed as the most probable distribution of maximum wave heights from this event while the absolute maximum wave heights can be seen as a worst case scenario. In the absolute maximum wave height case, non-zero wave heights are seen along the coastline of southern France and Sardinia. These 'on the fly' ensemble results could augment the warnings deduced from the traditional decision matrix or precomputed databases. Further, the absolute and mean maximum wave heights can be normalised to a traditional non-dimensional warning level scale, which in turn can be given to the relevant authorities \cite{Gailler2013}.

\subsection{North East Atlantic Tsunami}
The last major tsunami to effect the Atlantic basin is the Lisbon 1755 event. This tsunami was generated by a $\approx 8.5$ Mw earthquake which occurred on the morning of the $1^{st}$ November 1755 roughly 100km off the coast of Lisbon. The earthquake and subsequent tsunami caused a large number of fatalities and severe damage along the coasts of Portugal, Spain and Morocco. The effects of the tsunami were felt across the Atlantic basin, with historical reports of the tsunami waves arriving in the Caribbean, United Kingdom and Ireland \cite{OBrien2013}. There is still no general consensus on the exact earthquake parameters for the event. With the effect of this event on the Irish coastline in mind a 'reduced' faster than real time ensemble will be shown. This is considered a 'reduced' ensemble as only six different earthquake sources were selected from the literature and only a section of the north east Atlantic was chosen. The initial sea surface was again generated using the Okada model \cite{Okada1985}. The resolution of the grid is 2 arc minutes and contains $\approx$ 610,000 cells. Each individual simulation is run for 8 hours and  then the maximum wave heights at each grid point are outputted. The complete run time for the six different tsunami simulations is 93s on two Nvidia V100 GPUs. 

\begin{figure*}[h!]
	\centering
	\includegraphics[width = 16cm, height = 12cm]{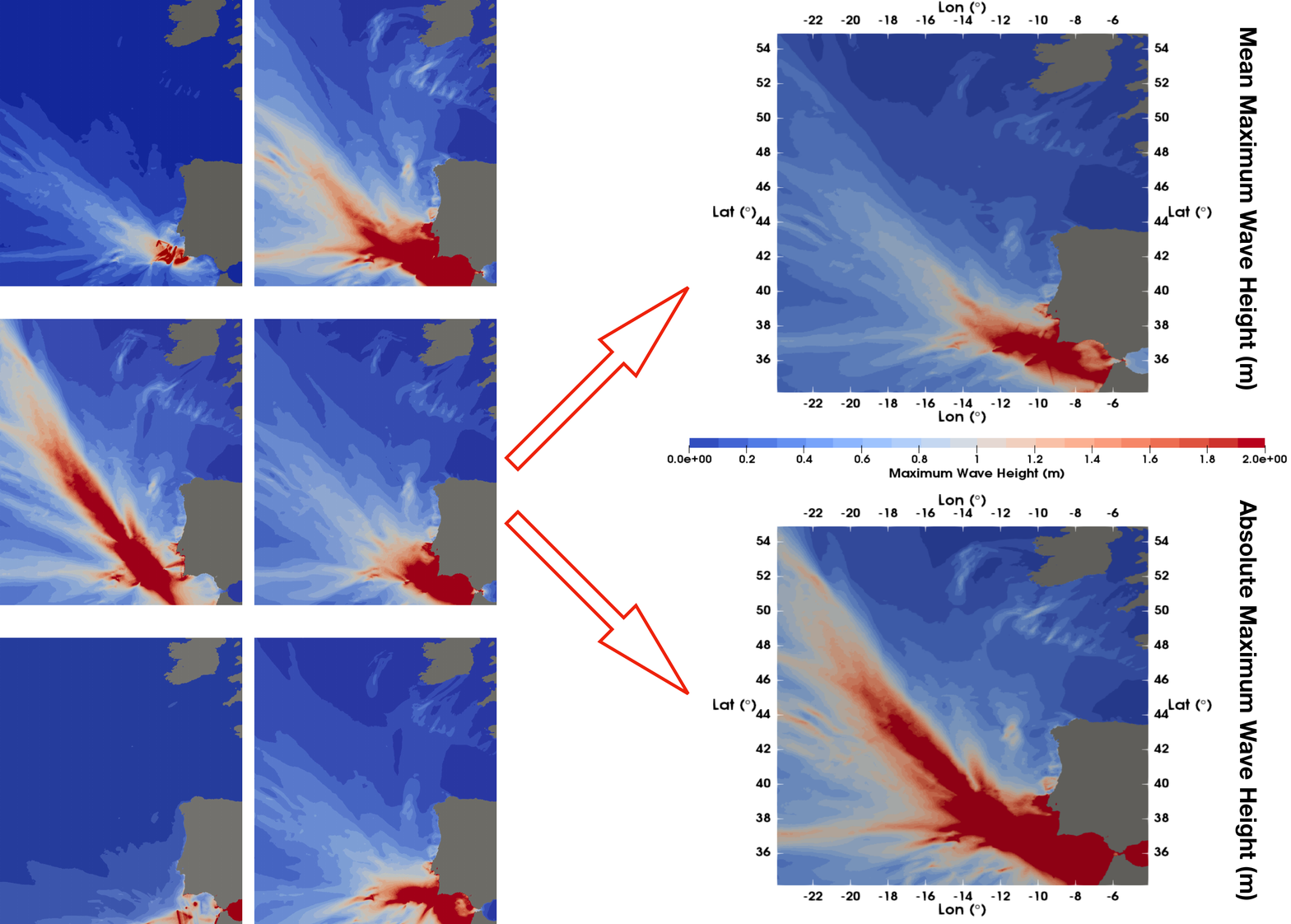}
	\caption[Med_warning]{Left Figure: Maximum wave heights from each of the six independent simulations. Right Figures: The combined mean maximum wave heights and absolute maximum wave heights.} 
	\label{atlantic_warning}
\end{figure*}

The plots of the maximum wave heights all highlight the impact on the local coastlines of Portugal, Spain and Morocco. They however also show a substantial variation in the directionality of the wave energy outside the gulf of Cadiz, with some sources showing the wave energy propagating towards the Irish Coastline. The absolute and mean maximum wave heights provide a rough estimate at a worst case and most likely scenario respectively. They are considered a rough estimate as the ensemble contains so few members.  To investigate the effect of this event on the Irish coastline the sources introduced in the above figure will be investigated further. 
\clearpage

\section{High Resolution Inundation Mapping}\label{Inundation}
\subsection{Most Impactful Earthquake Source}

From an Irish perspective the main tsunami threats lie in a Lisbon type event or a tsunami originating from a Mid-Atlantic submarine landslide. Extensive studies have been carried out on the collapse of the Rockall bank slide complex (located off the north-west coast of Ireland) and the subsequently generated tsunami \cite{Salmanidou2017}. However, this inundation mapping study focuses on earthquake generated tsunamis and from an Irish perspective the Lisbon 1755 event can be considered a worst case scenario. 

\begin{figure*}[h!]
	\centering
	\includegraphics[width = 20cm, height = 12cm]{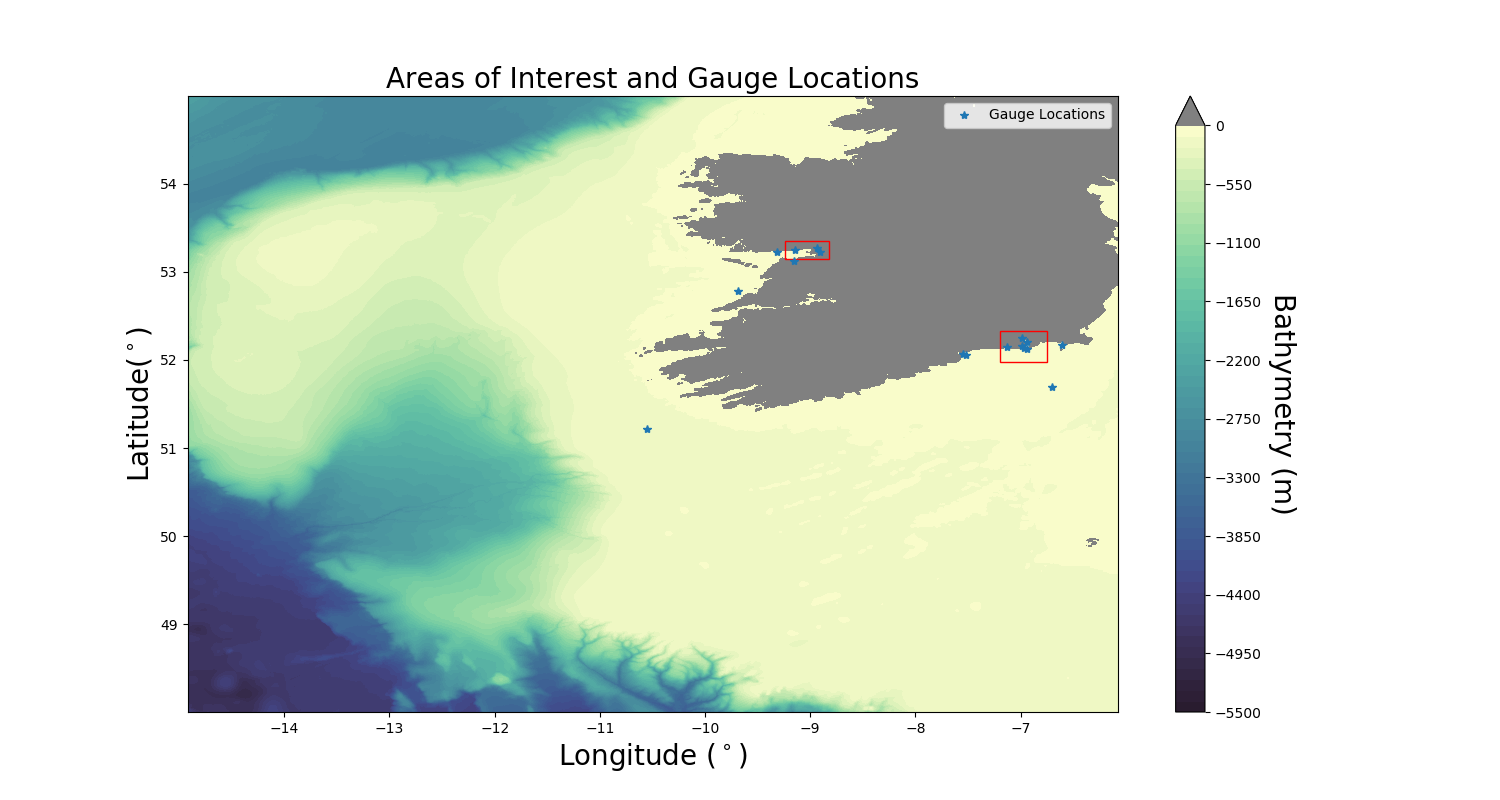}
	\caption[]{Zoom in on the bathymetry around the coast of Ireland, with the areas of interest; Galway Bay and Dunmore East highlighted by the red squares. The location of various wave gauges are highlighted by the blue stars.  \normalsize} 
	\label{Ireland_Bathy}
\end{figure*}

\begin{figure*}[h!]
	\centering
	\includegraphics[width = 18cm, height=12cm]{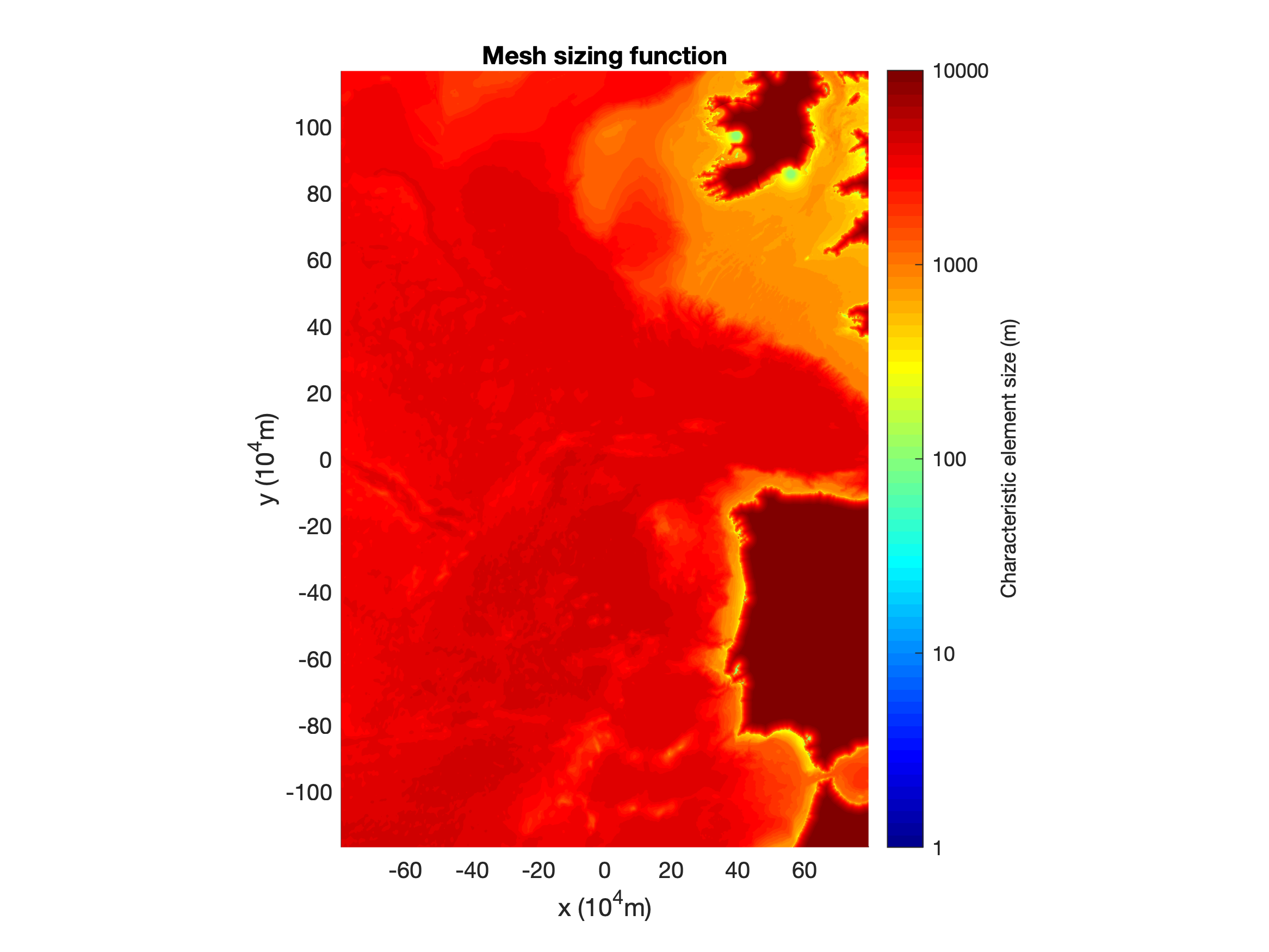}
	\caption{Non-uniform unstructured mesh which is refined near the coastlines and the areas of interest; Galway Bay and Dunmore East. The most impactful source on the Irish coastline is deduced from simulations with this mesh. } 
	\label{fig:mesh1}
\end{figure*}

\begin{figure}[h!]
        \begin{subfigure}[b]{0.475\textwidth}  
	\centering
	\includegraphics[width = 9cm, height = 6.1cm]{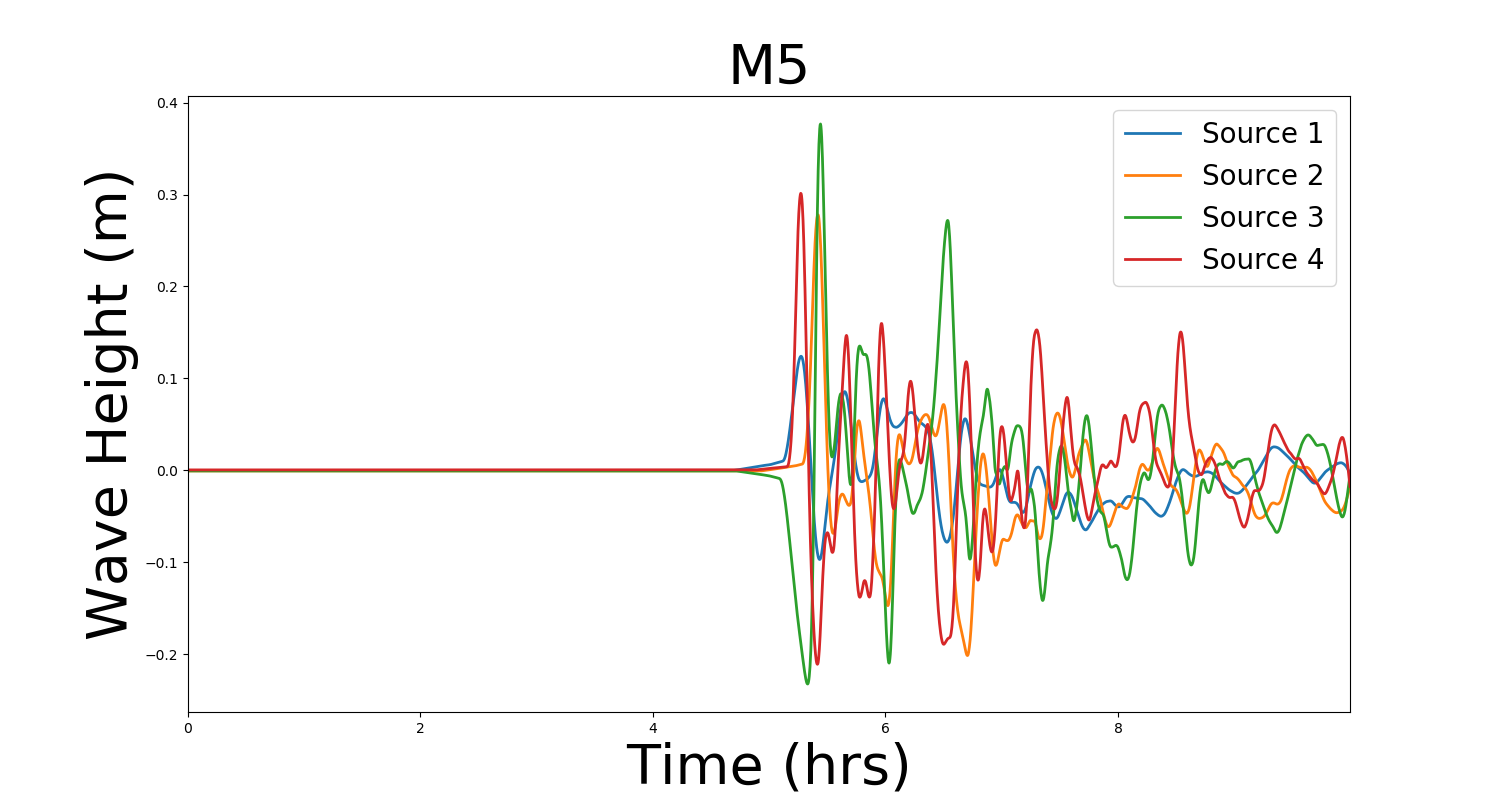}
	\end{subfigure}
	\quad
	\begin{subfigure}[b]{0.475\textwidth} 
	\centering
	\includegraphics[width = 9cm, height = 6.1cm]{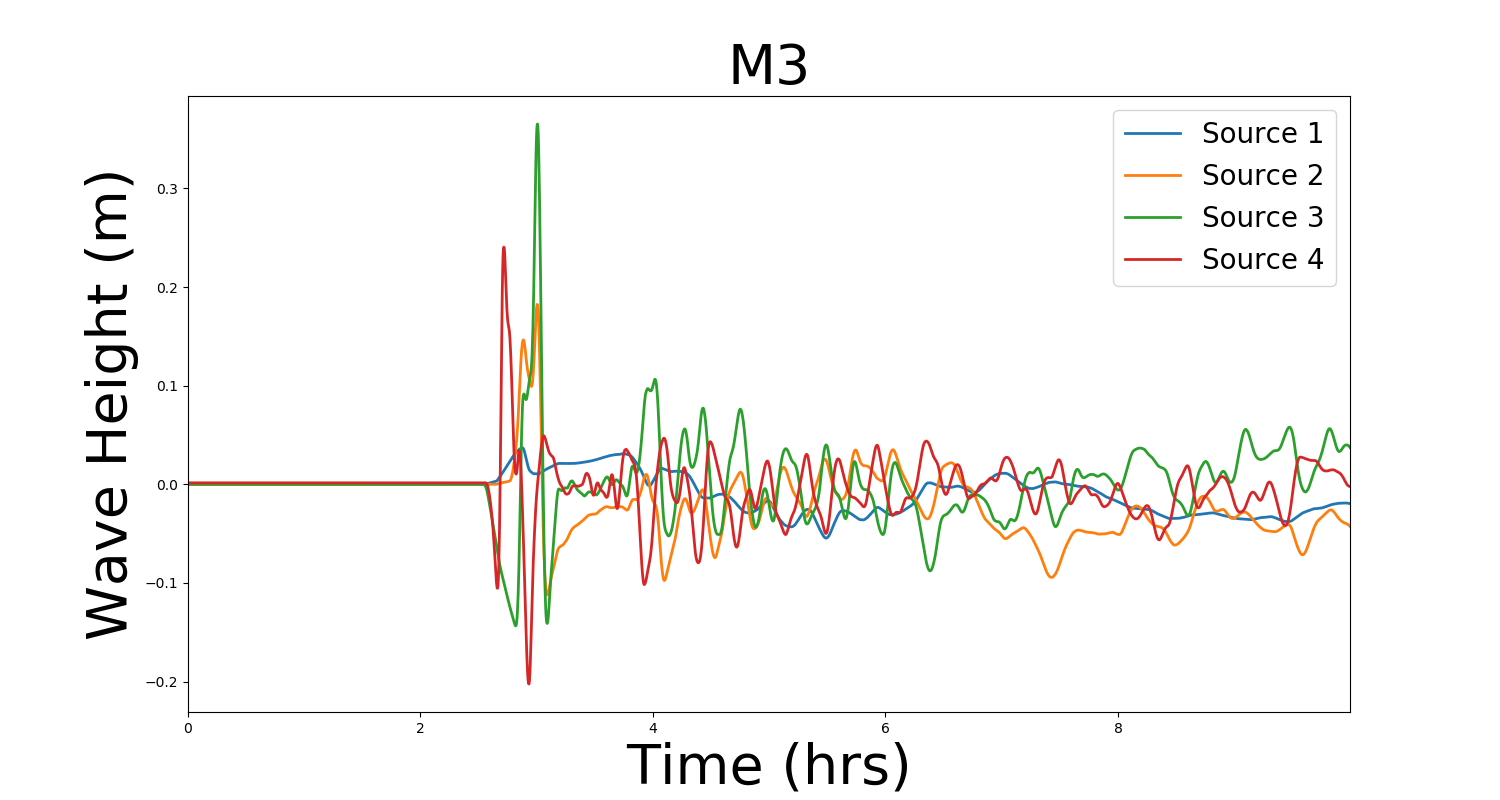}
	\end{subfigure}\vspace{-10pt}
	\vskip\baselineskip
	
	\begin{subfigure}[b]{0.475\textwidth} 
	\centering
	\includegraphics[width = 9cm, height = 6.1cm]{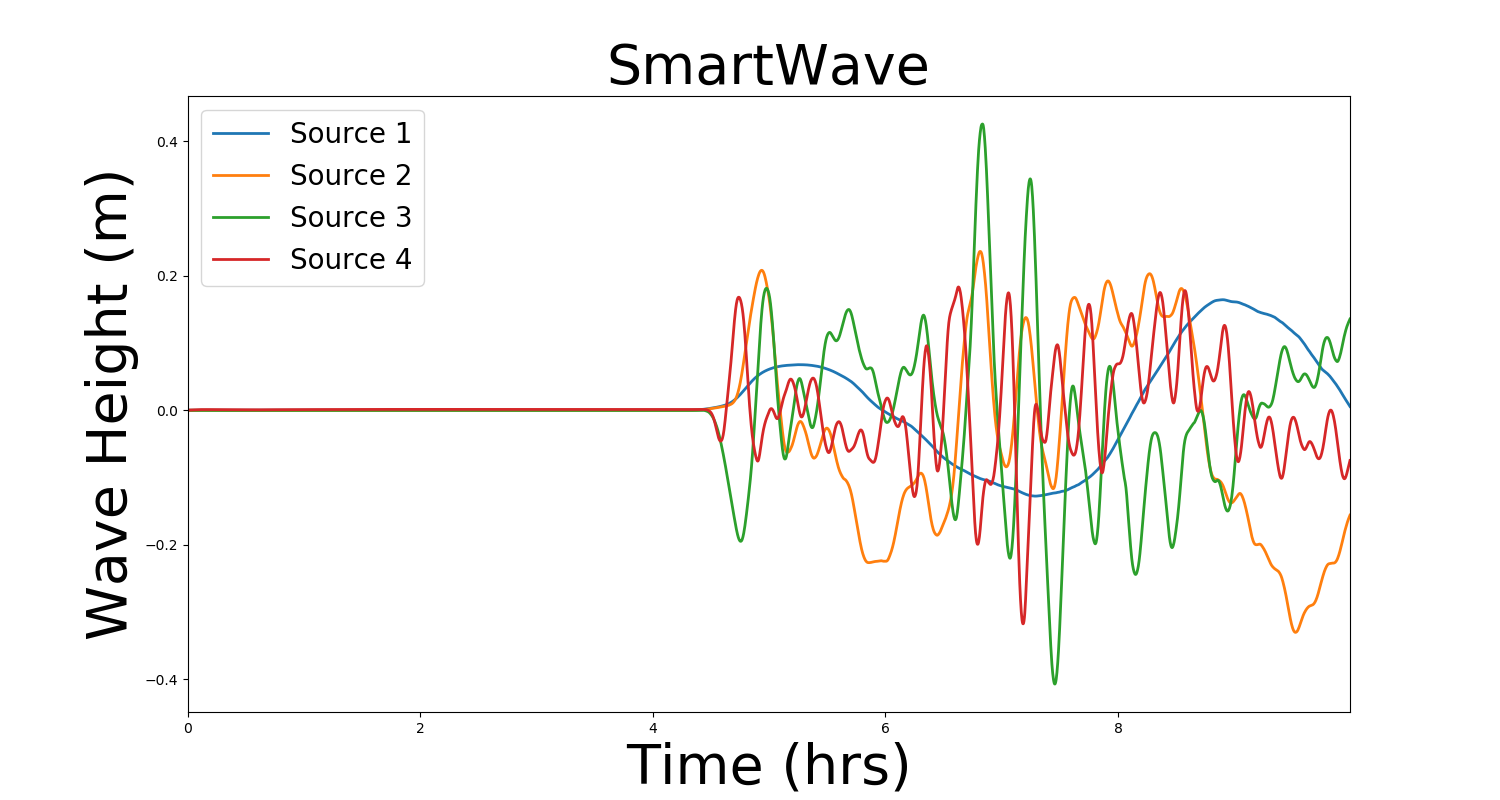}
	\end{subfigure}
	\quad
	\begin{subfigure}[b]{0.475\textwidth} 
	\centering
	\includegraphics[width = 9cm, height = 6.1cm]{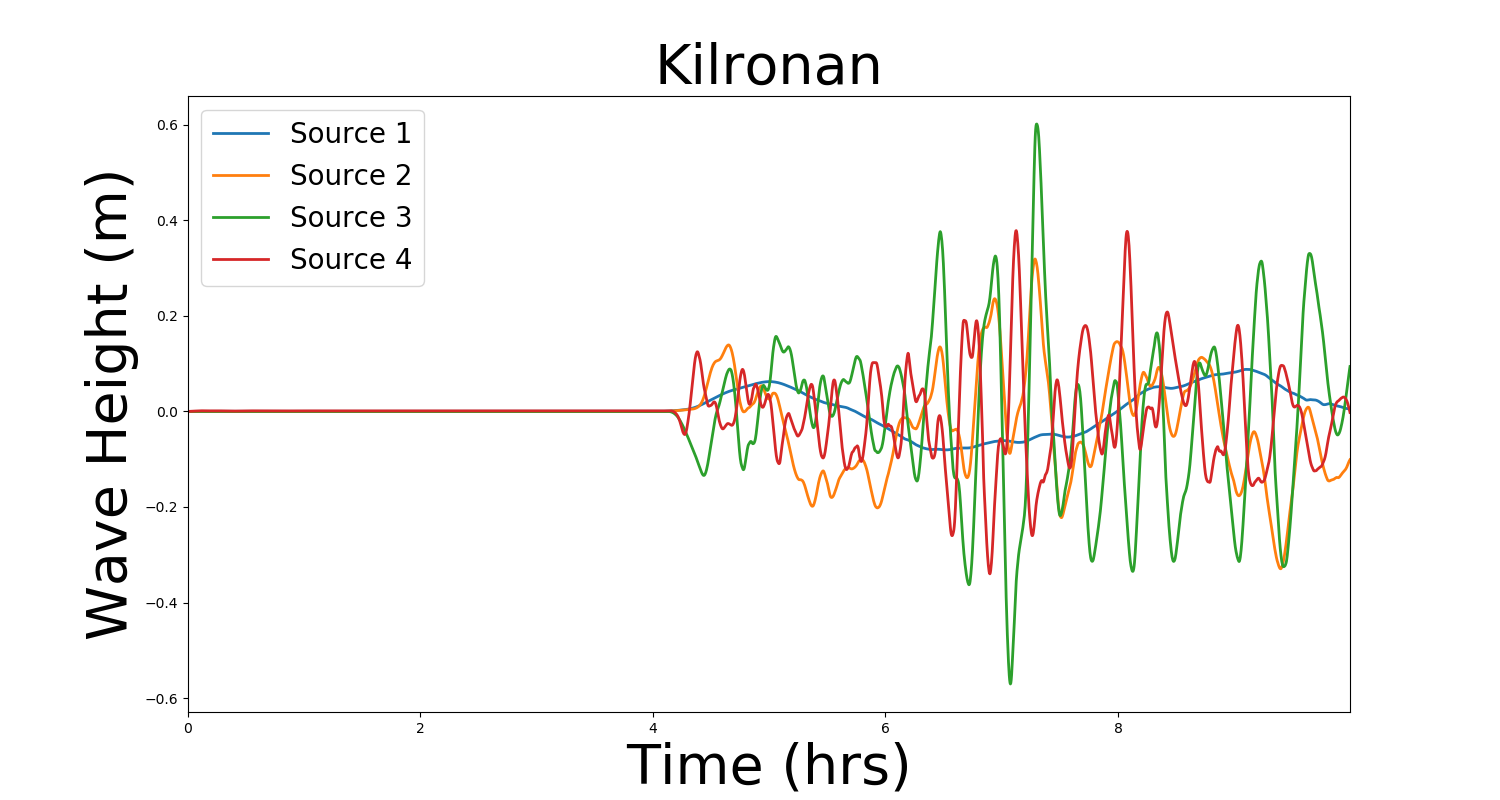}
	\end{subfigure}\vspace{-10pt}
	\vskip\baselineskip
	
        \begin{subfigure}[b]{0.475\textwidth}  
	\centering
	\includegraphics[width = 9cm, height = 6.1cm]{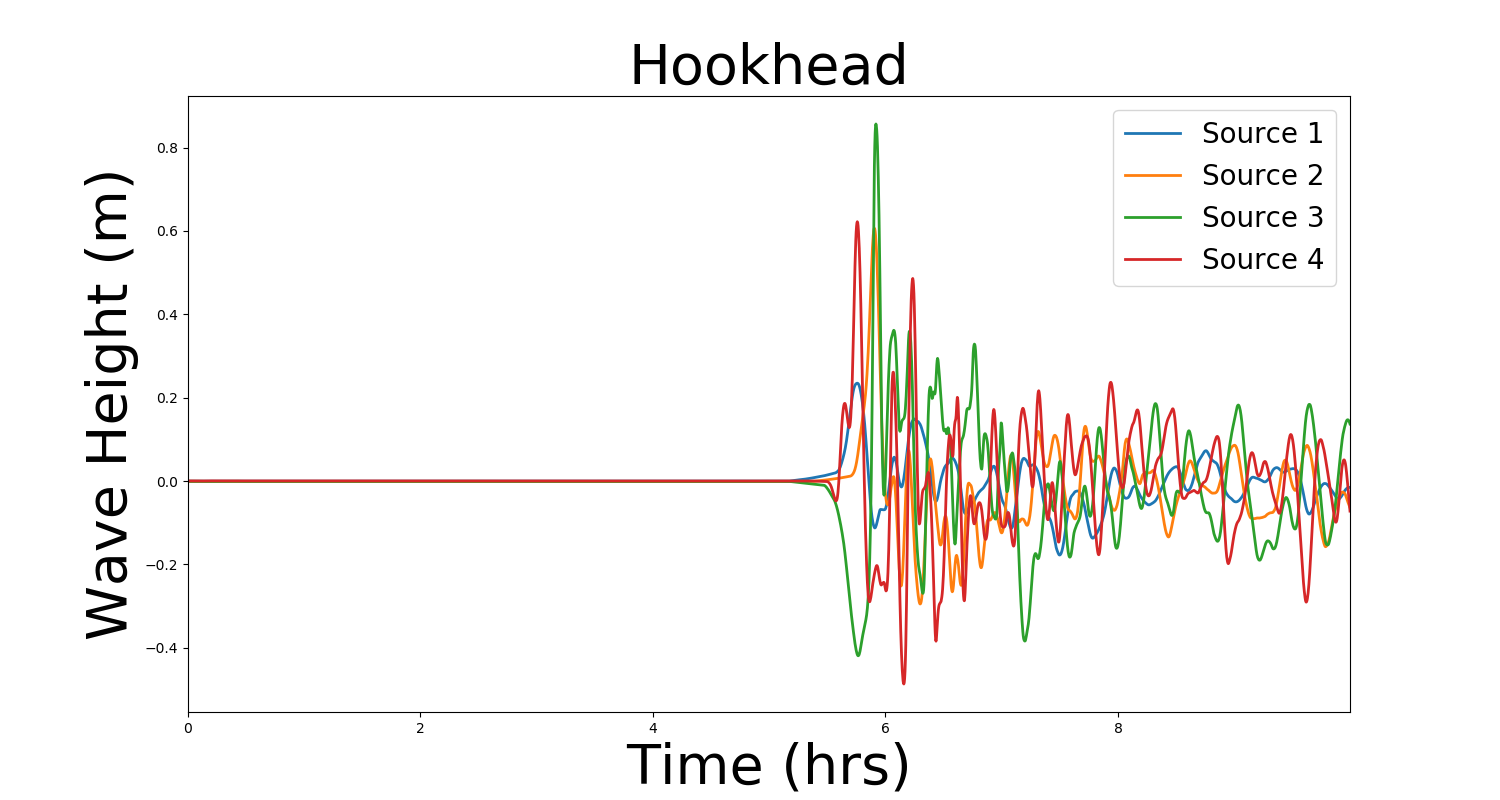}
	\end{subfigure}
	\quad
	\begin{subfigure}[b]{0.475\textwidth} 
	\centering
	\includegraphics[width = 9cm, height = 6.1cm]{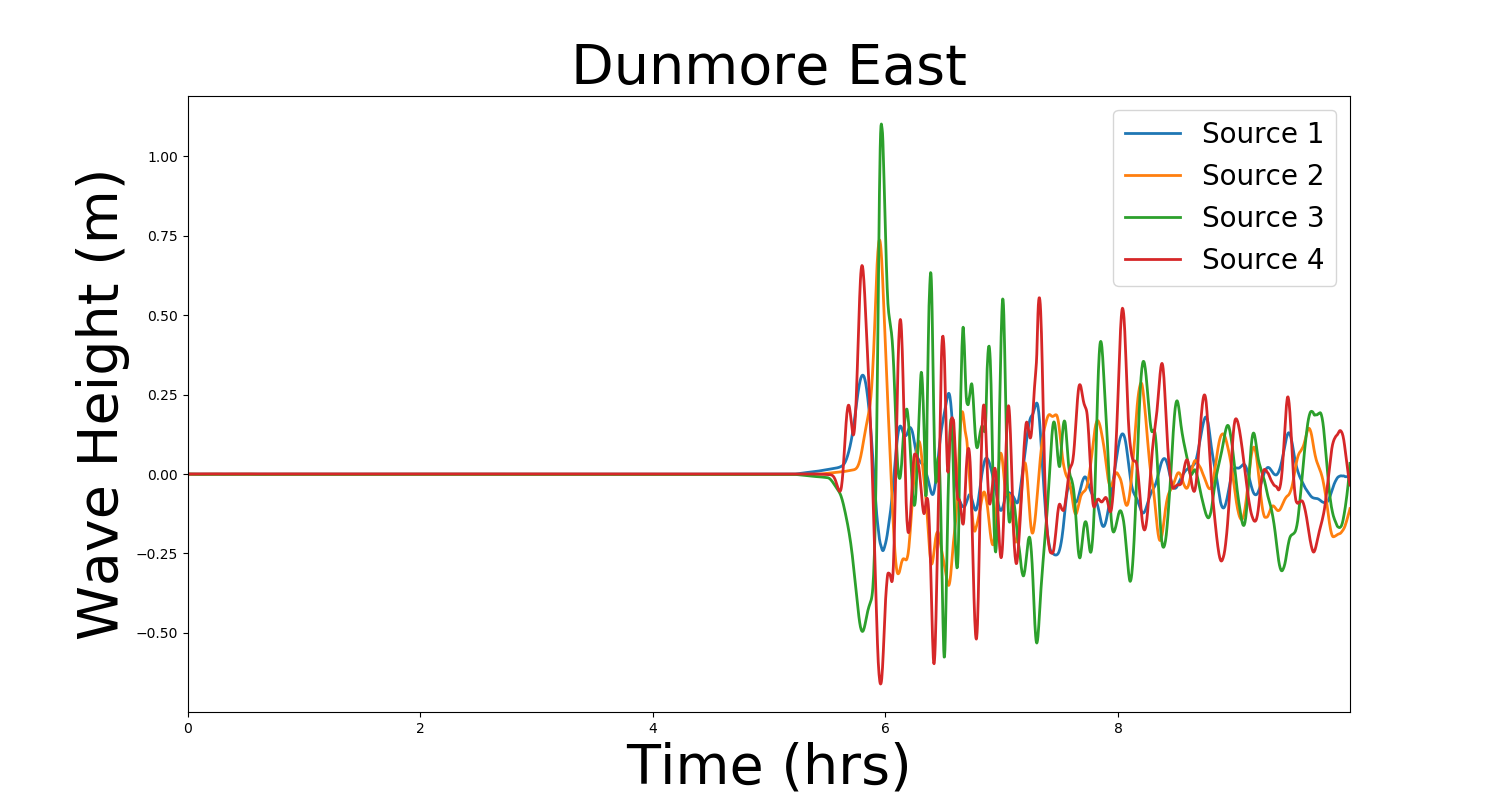}
	\end{subfigure}
	\caption{Wave gauge plots from selected locations around the coastline in Ireland. Top row: Wave surface elevations from the four different sources at wave buoys located off the coast of Ireland. Middle Row: Water surface elevations from the four different sources at gauges located in Galway Bay. Bottom Row: Water surface elevations from the four different sources at gauges located in the Dunmore East area. } 
	\label{fig:gauges}
\end{figure}

\begin{figure}[h!]
        \begin{subfigure}[b]{0.475\textwidth}  
	\centering
	\includegraphics[width = 10.5cm, height = 9cm]{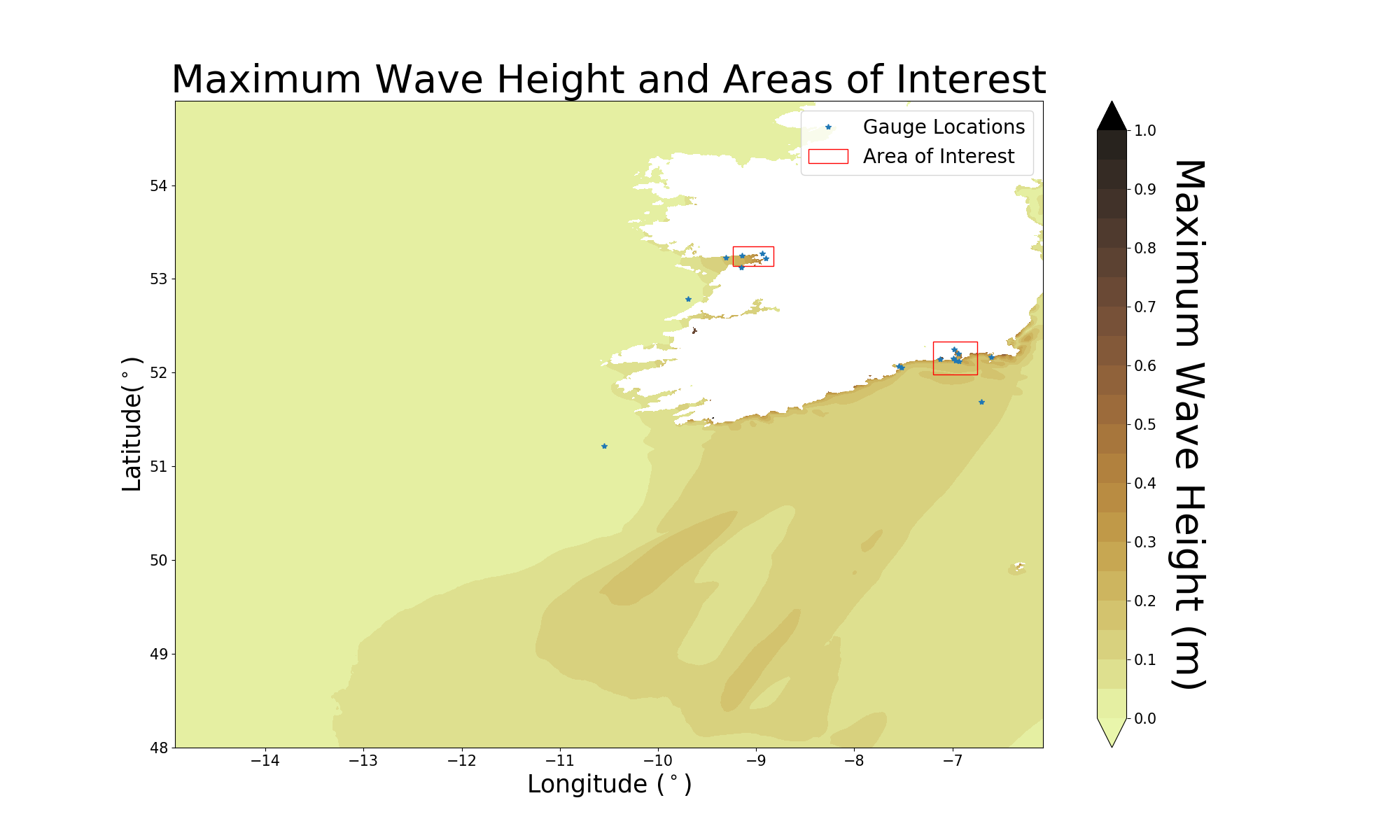}
	\end{subfigure}
	\quad
	\begin{subfigure}[b]{0.475\textwidth} 
	\centering
	\includegraphics[width = 10.5cm, height = 9cm]{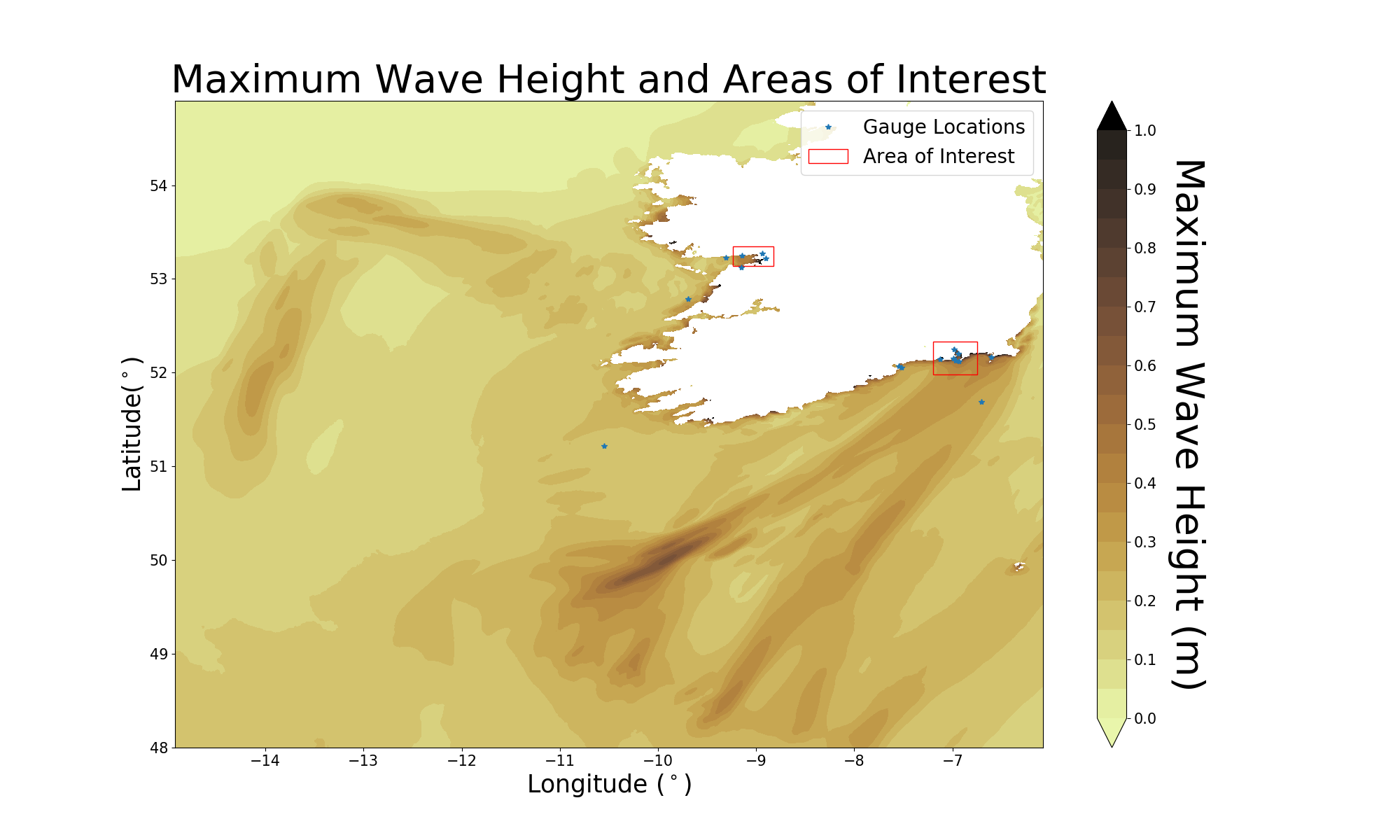}
	\end{subfigure}\vspace{-10pt}
	\vskip\baselineskip
	 \begin{subfigure}[b]{0.475\textwidth}  
	\centering
	\includegraphics[width = 10.5cm, height = 9cm]{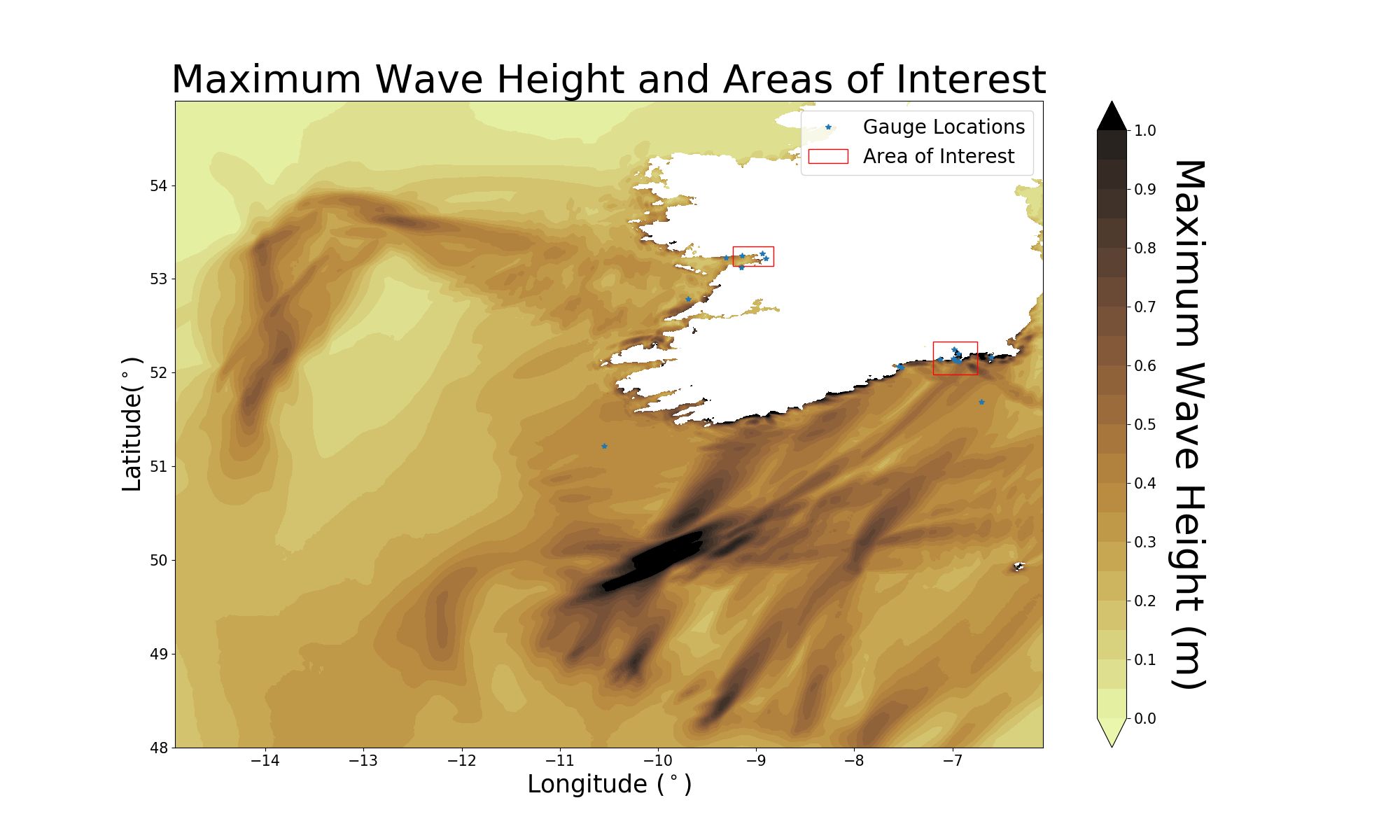}
	\end{subfigure}
	\quad
	\begin{subfigure}[b]{0.475\textwidth} 
	\centering
	\includegraphics[width = 10.5cm, height = 9cm]{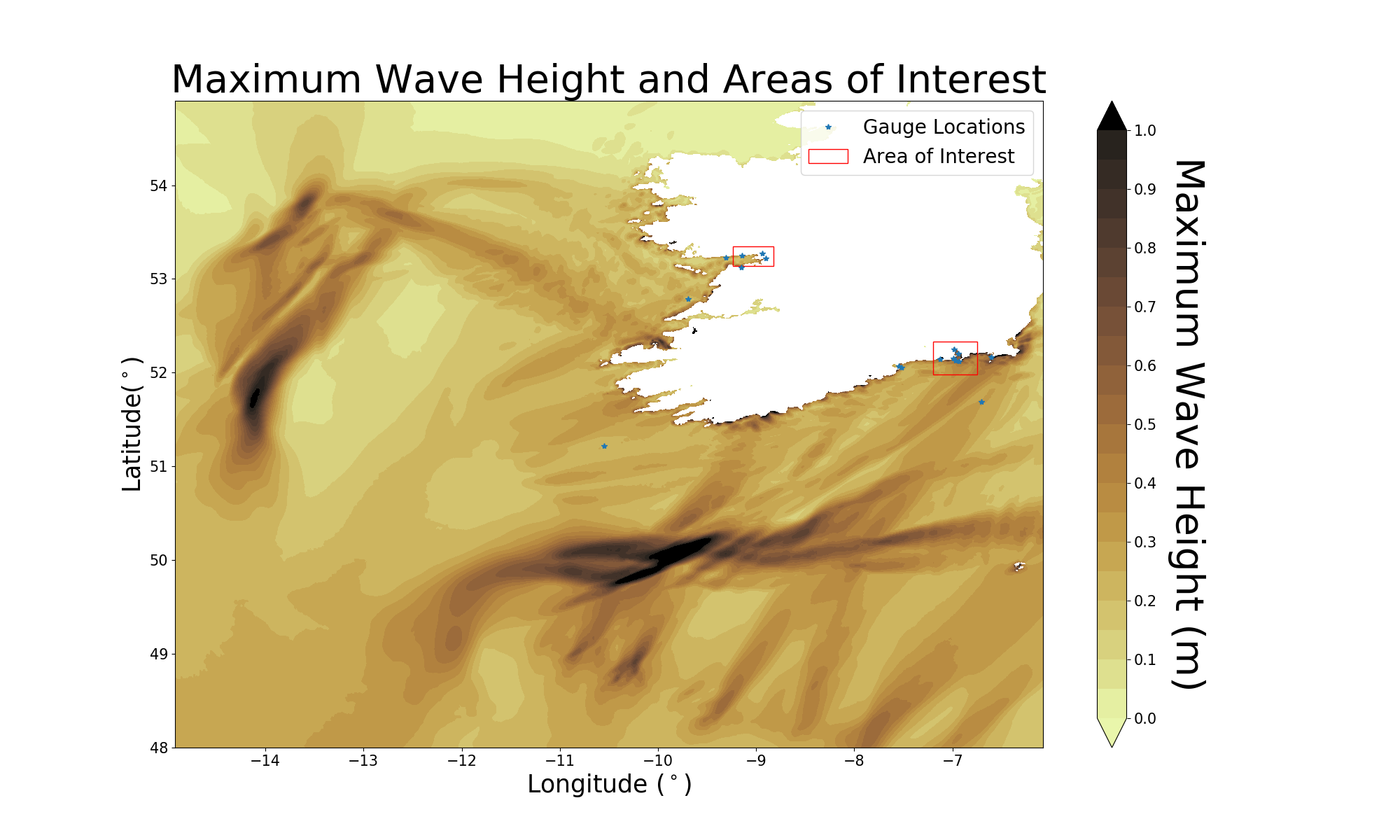}
	\end{subfigure}\vspace{-10pt}
	\vskip\baselineskip
	\caption{Zoom in on the maximum wave heights simulated for the four sources around the coast of Ireland. The numerical wave gauges and the areas of interest are highlighted in each plot. Top Left: Source 1, Top Right: Source 2, Bottom Left: Source 3 and Bottom Right: Source 4.} 
	\label{maxheights}
\end{figure}

Previous work on the effects of the Lisbon tsunami have highlighted the vulnerable sections of the Irish coastline to be the west (Galway Bay) and south east (Dunmore East) coastlines (areas highlighted in Fig \ref{Ireland_Bathy}).  The south-east area was chosen based on a previous study \cite{Defra2006} which highlighted that this area was prone to tsunami amplification. The west coast was selected based on tsunami simulations carried out by the French tsunami warning Centre (CENALT). The CENALT simulations showed that the tsunami wave is directed into Galway Bay and amplified because of refraction from the Porcupine Bank.
The reduced ensemble results (Fig \ref{atlantic_warning}) corroborate these findings, with the main tsunami energy impacting the west and south east coastline of Ireland. To build on the reduced ensemble, the sources which exhibited higher impact (maximum wave height) on the Irish coastline were chosen (Table \ref{tab:sources}). These sources were then re-run on a non-uniform mesh which is refined around the areas of interest (Fig \ref{fig:mesh1}).  The resultant maximum wave height plots (Fig \ref{maxheights}) and outputs at various wave gauges (Fig \ref{fig:gauges}) allowed the most impactful source to be deduced. This source was then used for the high resolution inundation mapping (section \ref{mapping}) and can represent a reasonable worst case scenario.

\clearpage

\underline{\textbf{Description of Source Parameters}}
\begin{center}
 \begin{tabular}{||c c c c c c c||} 
 \hline
 Source & Length(km) & Width(km) & Dip (\degree) & Strike (\degree) & Slip (m) & Rake (\degree) \\ [0.5ex] 
 \hline\hline
 1 (MPTF) & 105 & 55 & 24 & 21.7 & 20 & 90 \\ 
 1 (GB) & 96 & 55 & 45 & 70 & 20 & 90 \\
 \hline\hline
 2 (HF) & 165 & 70 & 35 & 42.1 & 15 & 90 \\
 \hline\hline
 3 (HF) & 270 & 45 & 70 & 44.2 & 22.5 & 90 \\
 \hline\hline
  4 (GorB) & 200 & 80 & 40 & 50 & 13.1 & 90 \\
  \hline
\end{tabular}\label{tab:sources}
\end{center}
\begin{itemize}

\item \underline{Source 1:}
A composite source which was first proposed by Baptista et al {\cite{Baptista2003}}. It involves the simultaneous rupture of the Guadalquivir Bank Fault (GB) and Marqu\^es de Pombal Thrust Fault (MPTF).\newline

\item \underline{Source 2:}
The second source involves the rupture of the Horseshoe Fault (HF) and has been purposed by Baptista et al. {\cite{Baptista2011}} for a tsunami inundation study of the Lisbon downtown area. The source was scaled in order to obtain rupture parameters comparable to those outlined by Solares and Arroyo {\cite{MartinezSolares2004}}.  \newline

\item \underline{Source 3:}
The third source has been described by BRGM (Bureau de recherches g\'eologiques et mini\`eres) in a pilot study for the TANDEM project (\url{http://www-tandem.cea.fr}) and also involves the HF. \newline

\item \underline{Source 4:}
A source purposed by \cite{Grandin2007} that involves the rupture of the Gorringe Banks (GorB).  \newline
\end{itemize}

It can be seen from the  maximum wave heights (Fig \ref{maxheights}) that the overall wave energy directionality of the tsunami is consistent across sources ($2-4$). Source $1$ fails to produce signs of refraction of the wave by the Porcupine Bank at the wave height scale used  (Fig \ref{maxheights}). It is interesting to note the presence of wave amplification as it propagates onto the continental shelf. This phenomenon has been observed elsewhere and has been reproduced in experimental set ups \cite{Knowles2018}.  Based on the selected wave gauge plots (Fig \ref{fig:gauges}) and the maximum wave heights (Fig \ref{maxheights}), the candidate source was chosen to be source 3, which has been proposed by BRGM. It can be seen from the wave gauge plots (Fig \ref{fig:gauges}) that this source produces the maximum wave heights along the Irish coastline, with the maximum wave height of $\approx$1m being produced at Dunmore East. This source will now be used for the high resolution inundation mapping simulation.

It should be noted that from these results the south coastline was also highlighted for its vulnerability to tsunami impact. Unfortunately, high resolution topography data was unavailable to the authors at the time that this pilot inundation study was carried out.   

\clearpage

\subsection{High Resolution Meshes}
High resolution bathymetry/topography grids ($\approx$ 10m) for the Galway Bay and Dunmore East areas were used in the generation of the unstructured meshes for Volna-OP2. The topography for the grids was obtained from raw LiDAR data. In order to ensure that the bathymetry and LiDAR data was continuous, all the LiDAR data was referenced to WGS84 (a global coordinate system) and then locally vertically referenced to OD Malin. Two separate high resolution (25km-10m) meshes for the Galway and Dunmore East coastlines were then generated.

\begin{figure*}[h!]
        \begin{subfigure}[b]{0.475\textwidth}  
	\centering
	\includegraphics[width = 10cm, height = 8cm]{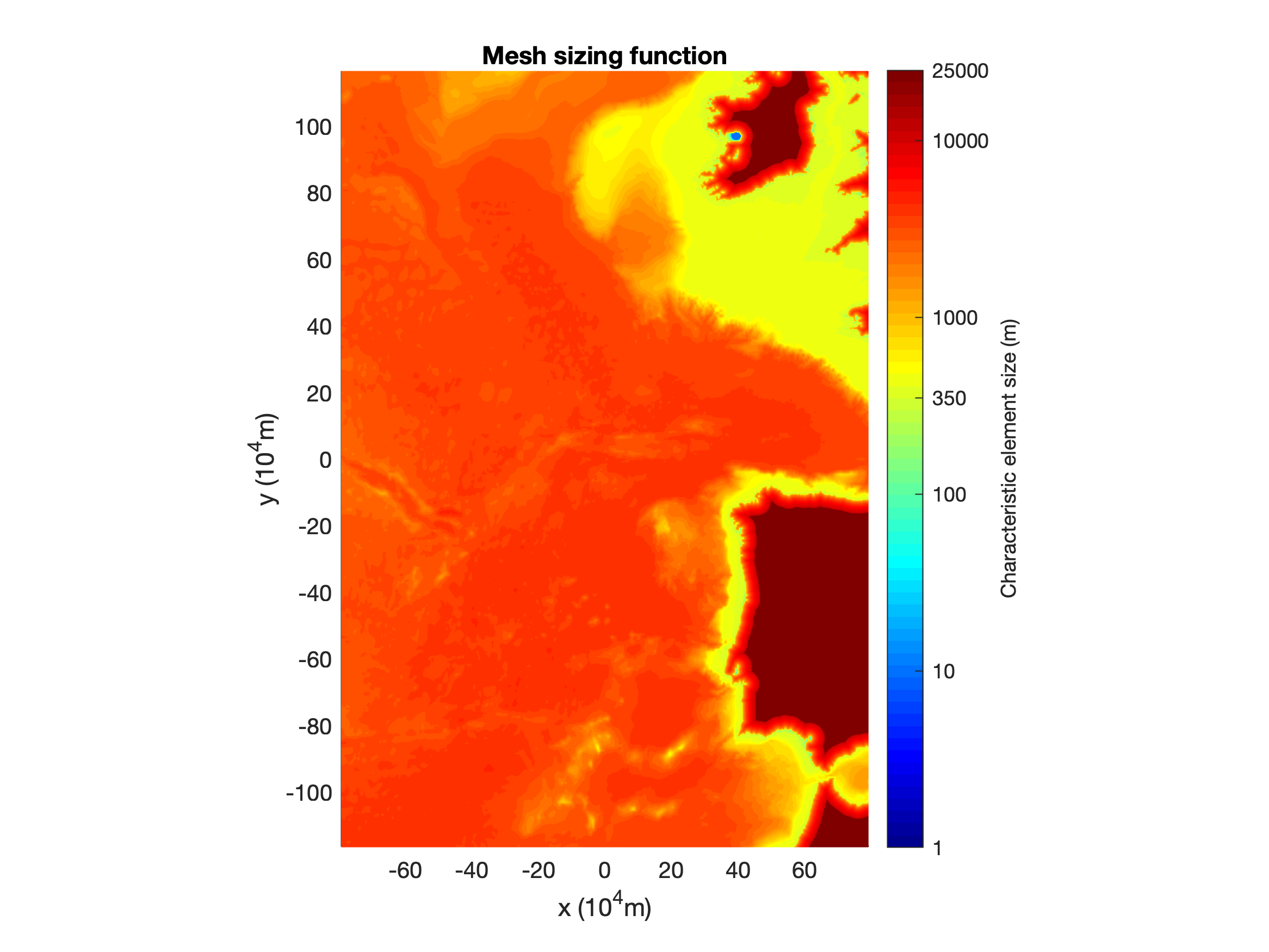}
	\end{subfigure}
	\begin{subfigure}[b]{0.475\textwidth} 
	\centering
	\includegraphics[width = 10cm, height = 8cm]{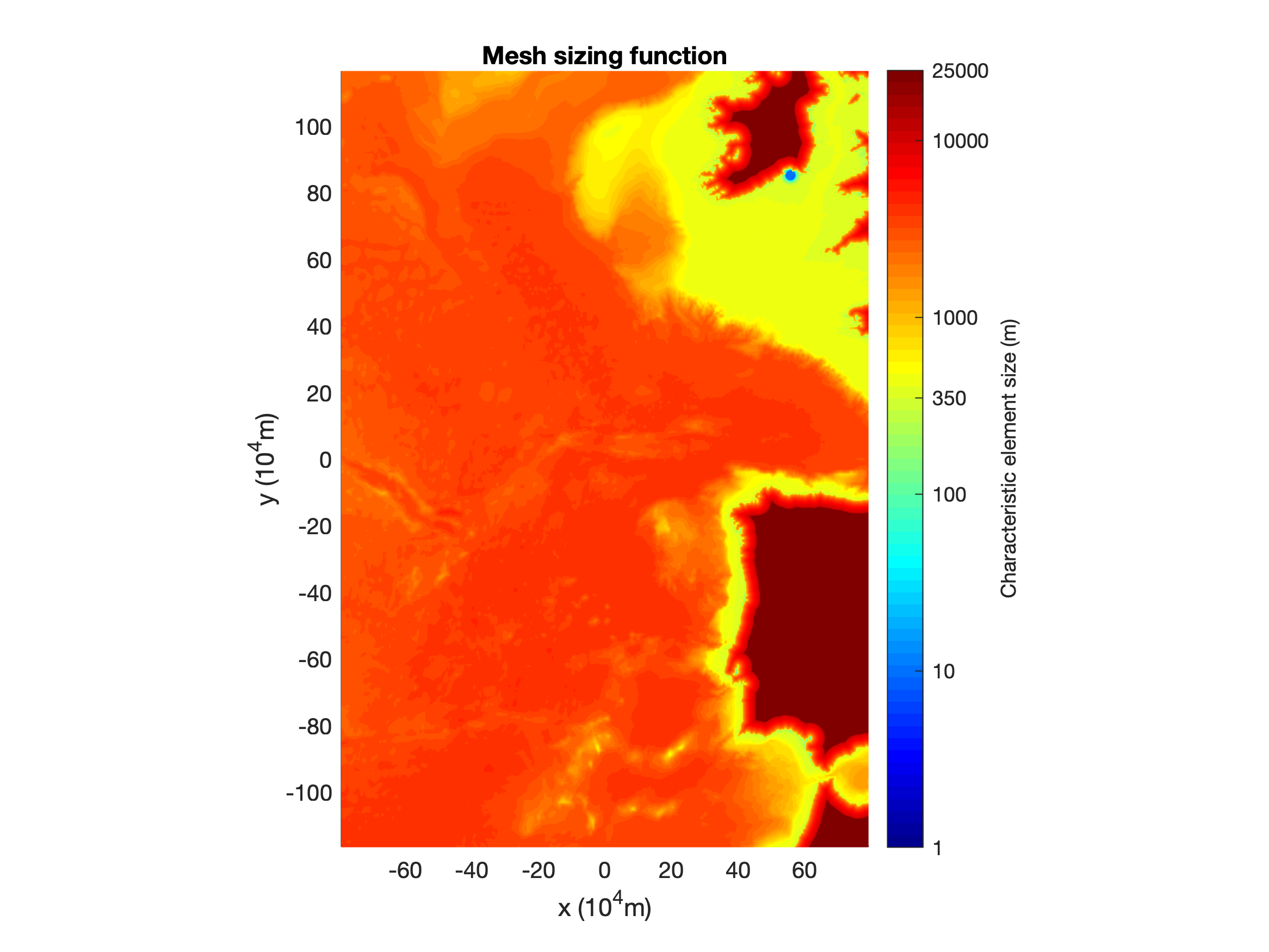}
	\end{subfigure}
	\caption{Size of mesh triangles in the unstructured mesh, Left: Galway Bay (25km-10m) and Right: Dunmore East (25km-10m). The differences can be seen in the concentration of 10 m resolution triangles surrounding Galway Bay and  Dunmore East in the left and right plots.  \normalsize} \label{fig:mesh2}
\end{figure*}

The nonuniform unstructured meshes (Fig \ref{fig:mesh1} and Fig\ref{fig:mesh2}) were generated using a customised mesh sizing function and the GMSH software \cite{Geuzaine2009}. The mesh sizing function spilts the domain into three separate regions: offshore, onshore and port region (area of interest). In the offshore region the mesh size is calculated based on the bathymetry value $b(x,y)$. Onshore cell sizes are dependent on the distance to the coastline while in the area of interest a fixed cell size is used. Full details on the mesh sizing function are given in \cite{Gopinathan2020}. 

\begin{figure*}[h!]
        \begin{subfigure}[b]{0.475\textwidth}  
	\centering
	\includegraphics[width = 10cm, height = 8cm]{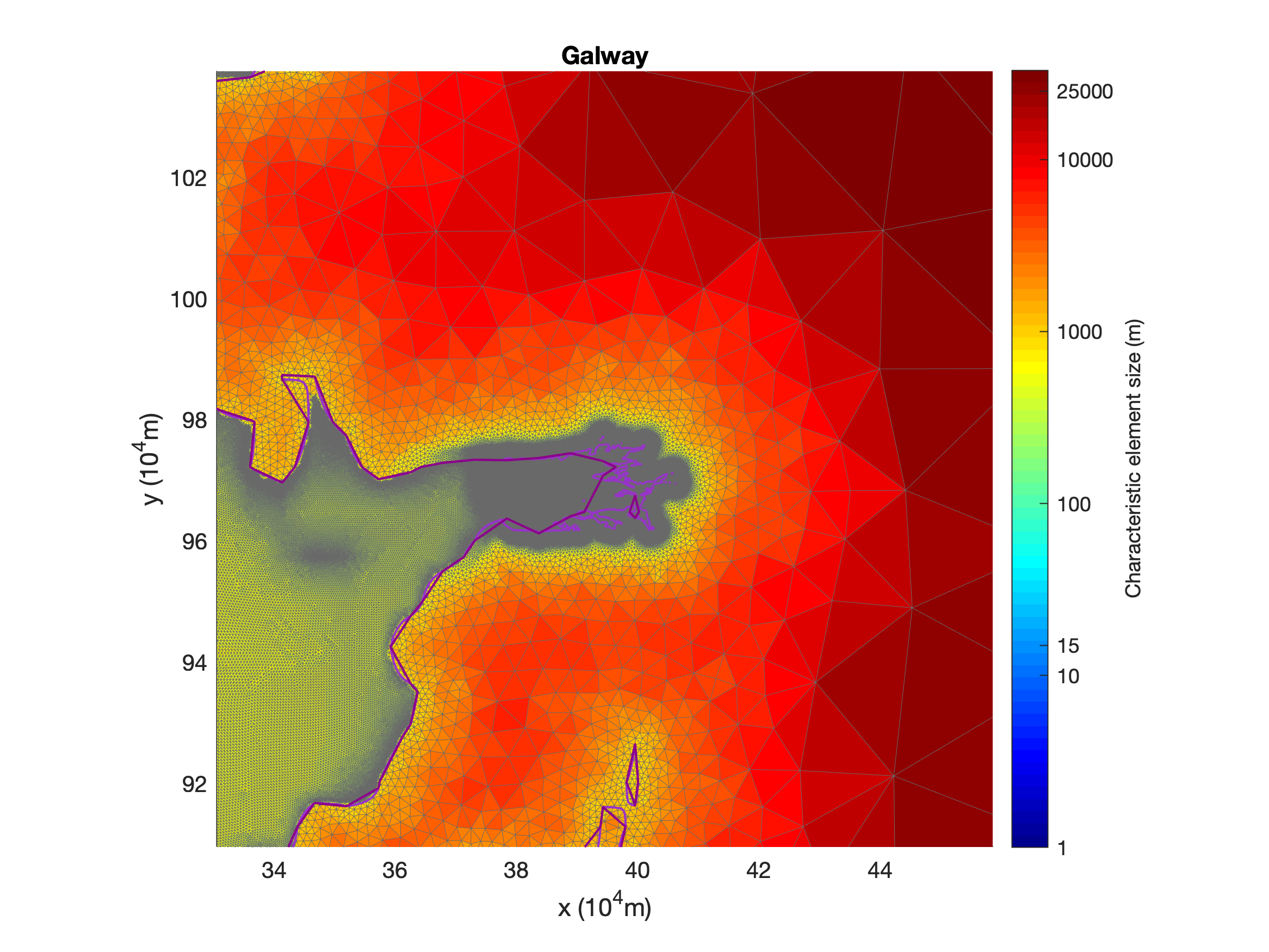}
	\end{subfigure}
	\begin{subfigure}[b]{0.475\textwidth} 
	\centering
	\includegraphics[width = 10cm, height = 8cm]{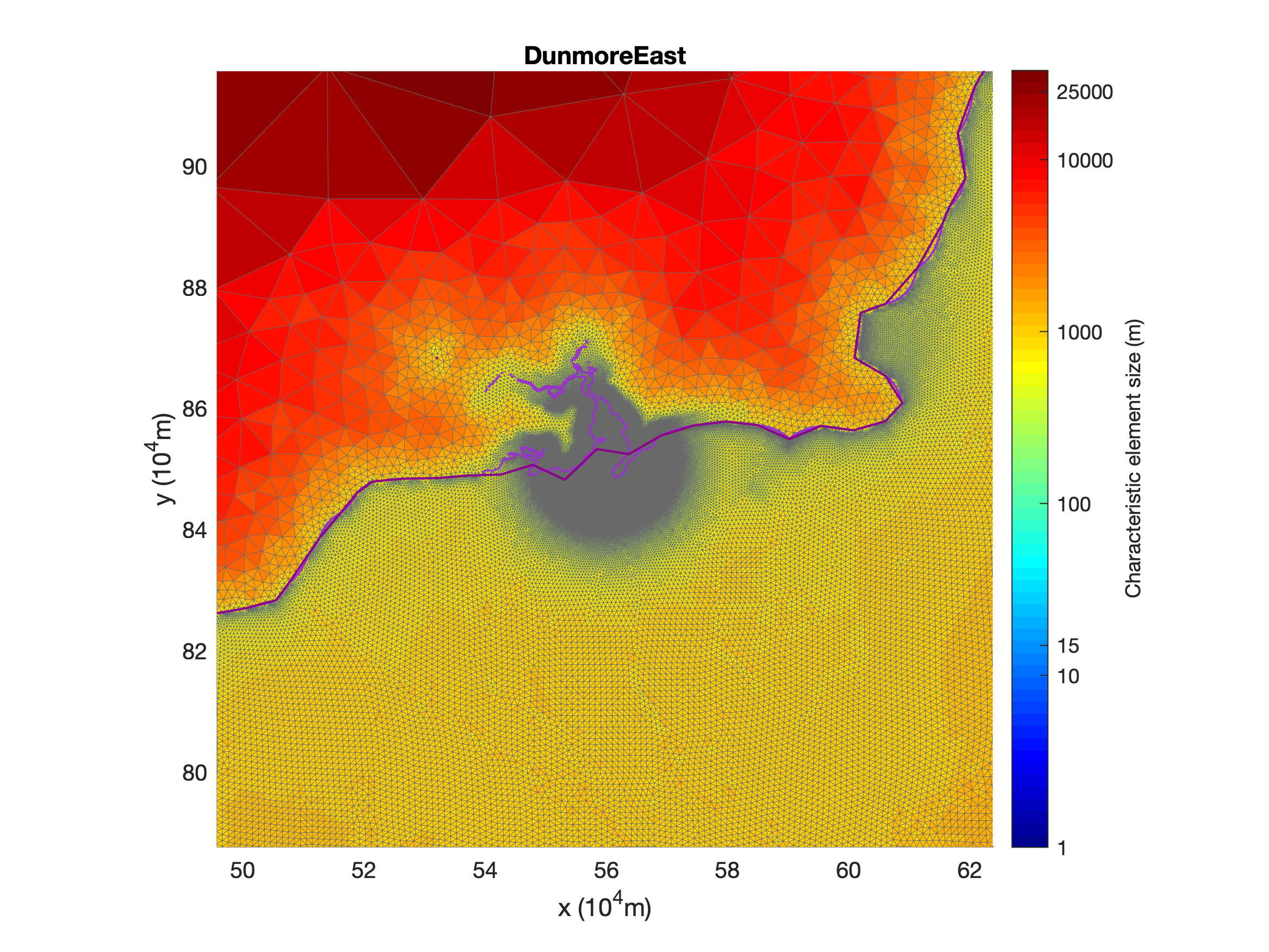}
	\end{subfigure}
	\caption{Zoom in on the high resolution mesh generated for, Left: Galway Bay and Right: Dunmore East.} 
\end{figure*}

\subsection{Inundation Maps}\label{mapping}
The following inundation maps highlight land areas which were inundated during the high resolution simulations ($\approx$ 10m resolution). The grey points indicate cells which were inundated during the simulation. The maximum run-up height was calculated to be $\approx$ 3.4m for both the Dunmore East and Galway Bay areas. It should be noted that the simulations with these highly non-uniform meshes were extremely dependent on the Courant-Friedrich-Levant (CFL) parameter used. Too high of a CFL condition and the numerical simulation became unstable, while too low of a value and the tsunami signal became diffused and dispersed. Thus, multiple preliminary runs were needed to optimise for this parameter.

\subsubsection{Galway Bay}
\begin{figure*}[h!]
	\centering
	\includegraphics[width = 16cm, height = 14cm]{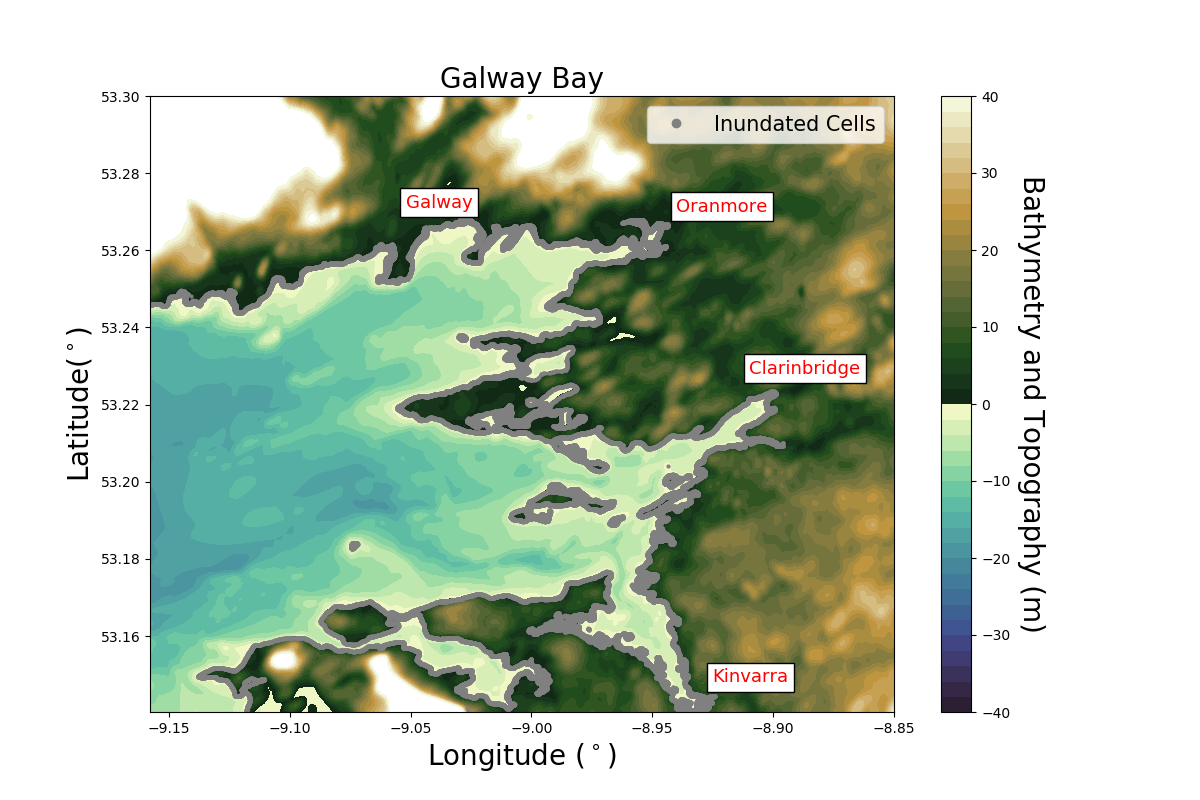}
	\caption[Inundation maps for Galway Bay (Volna-OP2).]{Inundation maps for Galway Bay: the grey areas on the map are land values which were inundated during the simulation. \normalsize} 
	\label{Inundation_galway_volna}
\end{figure*}

Figure (\ref{Inundation_galway_volna}) highlights the areas of inundation for the Galway Bay region. It can be seen that the model predicts inundation at Kinvarra, Clarinbridge and Oranmore (marked on map).  Further it predicts little inundation at Galway itself,  the lack of high resolution bathymetry/topography in the Galway city area (Mutton Island) could be responsible for this. 

\clearpage

\subsubsection{Dunmore East}
\begin{figure*}[h!]
	\centering
	\includegraphics[width = 16cm, height = 14cm]{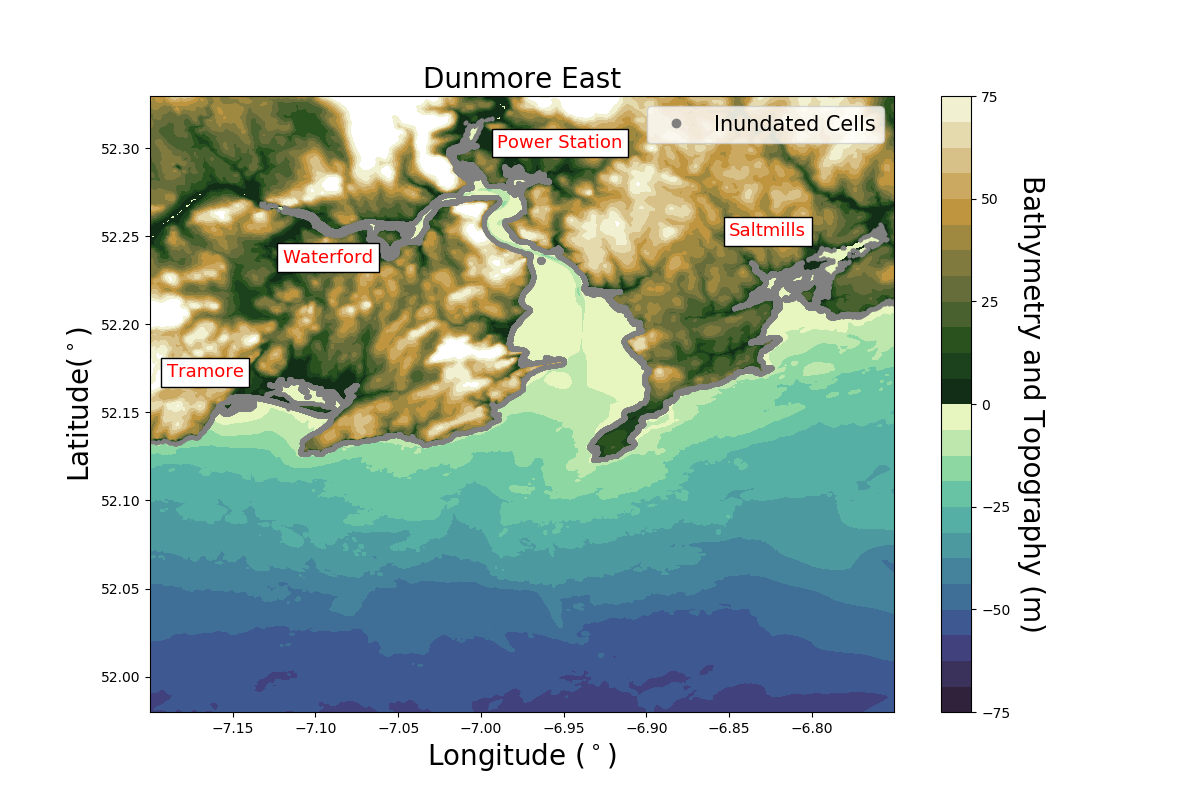}
	\caption[Inundation maps for Dunmore East (Volna-OP2).]{Inundation maps for Dunmore East: the grey areas on the map are land values which were inundated during the simulation. \normalsize} 
	\label{Inundation_dunmore_volna}
\end{figure*}

Figure (\ref{Inundation_dunmore_volna}) highlights the areas of inundation for the Dunmore East region. The areas of inundation include Tramore, Saltmills and near the Great Island Power Station. It should be noted that the flow in the River Barrow was not modelled in these simulations. This river flow could lead to the formation of a bore and thus allow the tsunami wave to propagate further up the river. The maximum water height at the Waterford coastline is predicted to be $\approx$ 1.2m. The maximum run-up height was calculated to be $\approx$ 3.4m.

\section{Concluding remarks}
The massively parallelised Volna-OP2 code is shown here to be a candidate code which could be implemented into a Tsunami Warning Centre's workflow. It is shown to produce extremely rapid regional forecasts for the Mediterranean Sea. The 2003 Boumerdes earthquake was chosen to act as an example for this process. Most tsunami warning centres now have access to substantial pre-computed databases of potential earthquake sources (1,000s of unit sources) and resultant tsunami wave heights \cite{Gailler2013}. Once a tsunami is detected, any relevant sources in the database can become an input into an 'on the fly' ensemble with Volna-OP2 being the tsunami code kernel. The resultant absolute maximum and mean wave height plots can augment the results from the pre-computed database.  The results presented here for the 'on the fly' ensembles are acknowledged to be preliminary and major effort in developing efficient workflows around the code is needed. However, it is clear that Volna-OP2 has promising potential when it comes to this respect. 

The codes capability at producing high resolution inundation studies is explored in section (\ref{Inundation}). This pilot inundation study highlights recent efforts to increase the tsunami hazard assessment for the Irish coastline. As stated, the main tsunami hazard for Ireland lies with a repeated Lisbon 1755 type event or a mid-Atlantic submarine landslide. The inundation study presented here assumes the Lisbon 1755 event as a worst case scenario and examines the potential inundation of sections of the Irish coastline from such a scenario. The impacts of various earthquake sources proposed in the literature are explored and the one which produced the highest maximum wave heights for the Irish coastline and at various gauge points was chosen for the inundation study. Utilising non-uniform unstructured meshes a high resolution simulation of the areas of interest was achieved with Volna-OP2. It was found that these high resolution inundation simulations were extremely dependent on the CFL condition, and thus required multiple preliminary runs to optimise for. 

The tsunami hazard assessment of the Irish coastline is considered a pilot study on the extent of inundation for the most vulnerable sections of the Irish coastline from a Lisbon type event. The results shown corroborate previous studies on what sections of coastline (the west and south east) are particularly vulnerable. The assessment found that inundation was constrained to low-lying areas in the Dunmore East and Galway Bay regions, with a maximum run-up of $\approx 3.4$m for both regions. The depth of water above the underlying topography (inundation depth) reached 2m in parts. This level of inundation could pose a risk to various infrastructures (Great Island Power Station) and members of the public located near the shoreline. To date the above pilot study provides the best estimate of assessing the hazard associated with a Lisbon type tsunami event on the Irish coastline. However, it is acknowledged that to provide a full hazard assessment of the Irish coastline, some further sites (south coast) and factors should be included, such as tidal forcing, 'street level' inundation and the effect of wave dispersion. Thus, the inundation plots produced should only be used as a guidance on the potential tsunami hazard of an area.

\vspace{6pt} 

\section*{Acknowledgments}
The authors wish to acknowledge the DJEI/DES/SFI/HEA Irish Centre for High-End Computing (ICHEC) for the provision of computational resources. Daniel Giles would like to acknowledge the support of the Government of Ireland Postgraduate Scholarship GOIPG/2017/68 from the Irish Research Council. The authors acknowledge funding and marine bathymetry and coastal LiDAR data from Geological Survey Ireland.


\bibliographystyle{alpha}
\bibliography{bib}

\newcommand{\etalchar}[1]{$^{#1}$}
\begin{thebibliography}{MMC{\etalchar{+}}04}

\bibitem[BDA{\etalchar{+}}04]{Bounif2004}
A.~Bounif, C.~Dorbath, A.~Ayadi, M.~Meghraoui, H.~Beldjoudi, N.~Laouami,
  M.~Frogneux, A.~Slimani, P.~J. Alasset, A.~Kharroubi, F.~Ousadou, M.~Chikh,
  A.~Harbi, S.~Larbes, and S.~Maouche.
\newblock {The 21 May 2003 Zemmouri (Algeria) earthquake Mw 6.8: Relocation and
  aftershock sequence analysis}.
\newblock {\em Geophysical Research Letters}, 31(19):2--5, 2004.

\bibitem[BG16]{Beck2016}
J.~Beck and S.~Guillas.
\newblock Sequential {D}esign with {M}utual {I}nformation for {C}omputer
  {E}xperiments ({MICE}): {E}mulation of a {T}sunami {M}odel.
\newblock {\em SIAM/ASA Journal on Uncertainty Quantification}, 4(1):739--766,
  2016.

\bibitem[BMCZ03]{Baptista2003}
M.~A. Baptista, J.~Miranda, F.~Chierici, and N.~Zitellini.
\newblock {New study of the 1755 earthquake source based on multi-channel
  seismic survey data and tsunami modeling}.
\newblock {\em Natural Hazards and Earth System Science}, 3:333--340, 2003.

\bibitem[BMOA11]{Baptista2011}
M.~A. Baptista, J.~Miranda, R.~Omira, and C.~Antunes.
\newblock {Potential inundation of Lisbon downtown by a 1755-like tsunami}.
\newblock {\em Natural Hazards and Earth System Science}, 11(12):3319--3326,
  2011.

\bibitem[DPD11]{Dutykh2011}
D.~Dutykh, R.~Poncet, and F.~Dias.
\newblock The {VOLNA} code for the numerical modeling of tsunami waves:
  Generation, propagation and inundation.
\newblock {\em European Journal of Mechanics, B/Fluids}, 30(6):598--615, 2011.

\bibitem[GBB{\etalchar{+}}07]{Grandin2007}
R.~Grandin, J.~F. Borges, M.~Bezzeghoud, B.~Caldeira, and F.~Carrilho.
\newblock {Simulations of strong ground motion in SW Iberia for the 1969
  February 28 (Ms = 8.0) and the 1755 November 1 (M$\sim$ 8.5) earthquakes -
  II. Strong ground motion simulations}.
\newblock {\em Geophysical Journal International}, 171(2):807--822, 2007.

\bibitem[GBB{\etalchar{+}}17]{Grezio2017}
A.~Grezio, A.~Babeyko, M.~A. Baptista, J.~Behrens, A.~Costa, G.~Davies, E.~L.
  Geist, S.~Glimsdal, F.~I. Gonz{\'{a}}lez, J.~Griffin, C.~B. Harbitz, R.~J.
  LeVeque, S.~Lorito, F.~L{\o}vholt, R.~Omira, R.~Mueller, C.and~Paris,
  T.~Parsons, J.~Polet, W.~Power, J.~Selva, M.~B. S{\o}rensen, and H.~K. Thio.
\newblock {Probabilistic Tsunami Hazard Analysis: Multiple Sources and Global
  Applications}.
\newblock {\em Reviews of Geophysics}, 55(4):1158--1198, 2017.

\bibitem[GHG20]{Gopinathan2020}
D.~Gopinathan, M.~Heidarzadeh, and S.~Guillas.
\newblock Probabilistic {Q}uantification of {T}sunami {C}urrents in {K}arachi
  {P}ort, {M}akran {S}ubduction {Z}one, using {S}tatistical {E}mulation.
\newblock {\em Earth and Space Science Open Archive}, 2020.

\bibitem[GHLH13]{Gailler2013}
A.~Gailler, H.~H{\'e}bert, A.~Loevenbruck, and B.~Hernandez.
\newblock Simulation systems for tsunami wave propagation forecasting within
  the {F}rench tsunami warning center.
\newblock {\em Natural Hazards and Earth System Sciences}, 13(10):2465, 2013.

\bibitem[GKS{\etalchar{+}}20]{Giles2020}
D.~Giles, E.~Kashdan, D.M Salmanidou, S.~Guillas, and F.~Dias.
\newblock Performance analysis of {V}olna-{OP2} -- massively parallel code for
  tsunami modelling, 2020.

\bibitem[GMS{\etalchar{+}}11]{Giles2011}
M.~B. Giles, G.~R. Mudalige, Z.~Sharif, G.~Markall, and P.~HJ. Kelly.
\newblock Performance analysis and optimization of the {OP2} framework on
  many-core architectures.
\newblock {\em The Computer Journal}, 55(2):168--180, 2011.

\bibitem[GP06]{Geist2006}
E.~L. Geist and Tom Parsons.
\newblock Probabilistic {A}nalysis of {T}sunami {H}azards*.
\newblock {\em Natural Hazards}, 37(3):277--314, Mar 2006.

\bibitem[GR09]{Geuzaine2009}
C.~Geuzaine and J-F. Remacle.
\newblock Gmsh: A 3-{D} finite element mesh generator with built-in pre- and
  post-processing facilities.
\newblock {\em International Journal for Numerical Methods in Engineering},
  79(11):1309--1331, 2009.

\bibitem[GSD{\etalchar{+}}18]{Guillas2018}
S.~Guillas, A.a Sarri, S.~J. Day, X.~Liu, and F.~Dias.
\newblock Functional emulation of high resolution tsunami modelling over
  {C}ascadia.
\newblock {\em Ann. Appl. Stat.}, 12(4):2023--2053, 12 2018.

\bibitem[GVR{\etalchar{+}}17]{Gopinathan2017}
D.~Gopinathan, M.~Venugopal, D.~Roy, K.~Rajendran, S.~Guillas, and F.~Dias.
\newblock Uncertainties in the 2004 {S}umatra{\textendash}{A}ndaman source
  through nonlinear stochastic inversion of tsunami waves.
\newblock {\em Proceedings of the Royal Society of London A: Mathematical,
  Physical and Engineering Sciences}, 473(2205), 2017.

\bibitem[{HR }06]{Defra2006}
{HR Wallingford, British Geological Survey, and Proudman Oceanographic
  Laboratory}.
\newblock {Tsunamis -- Assessing the Hazard for the UK and Irish Coasts}.
\newblock Technical report, {DEFRA (Department for Environment, Food and Rural
  Affairs), Flood Management Division, London, UK}, 2006.

\bibitem[KY18]{Knowles2018}
J.~Knowles and H.~Yeh.
\newblock {On shoaling of solitary waves}.
\newblock {\em Journal of Fluid Mechanics}, 848:1073--1097, 2018.

\bibitem[LG17]{Liu2017}
X.~Liu and S.~Guillas.
\newblock Dimension {R}eduction for {G}aussian {P}rocess {E}mulation: An
  {A}pplication to the {I}nfluence of {B}athymetry on {T}sunami {H}eights.
\newblock {\em SIAM/ASA Journal on Uncertainty Quantification}, 5(1):787--812,
  2017.

\bibitem[LLM{\etalchar{+}}19]{Lovholt2019}
F.~L{\o}vholt, S.~Lorito, J.~Mac{\'{I}}as, M.~Volpe, J.~Selva, and S.~Gibbons.
\newblock {Urgent tsunami computing}.
\newblock {\em Proceedings of UrgentHPC 2019: 1st International Workshop on HPC
  for Urgent Decision Making - Held in conjunction with SC 2019: The
  International Conference for High Performance Computing, Networking, Storage
  and Analysis}, 823844(823844):45--50, 2019.

\bibitem[MGR{\etalchar{+}}12]{Mudalige2012}
G.R. Mudalige, M.B. Giles, I.~Reguly, C.~Bertolli, and P.HJ. Kelly.
\newblock {OP2}: An active library framework for solving unstructured
  mesh-based applications on multi-core and many-core architectures.
\newblock In {\em Innovative Parallel Computing (InPar), 2012}, pages 1--12.
  IEEE, 2012.

\bibitem[ML04]{MartinezSolares2004}
J.M {Mart{\'{i}}nez Solares} and A.~{L{\'{o}}pez Arroyo}.
\newblock {The great historical 1755 earthquake. Effects and damage in Spain}.
\newblock {\em Journal of Seismology}, 8(1):275--294, 2004.

\bibitem[MMC{\etalchar{+}}04]{Meghraoui2004}
M.~Meghraoui, S.~Maouche, B.~Chemaa, Z.~Cakir, A.~Aoudia, A.~Harbi, P.~J.
  Alasset, A.~Ayadi, Y.~Bouhadad, and F.~Benhamouda.
\newblock {Coastal uplift and thrust faulting associated with the Mw = 6.8
  Zemmouri (Algeria) earthquake of 21 May, 2003}.
\newblock {\em Geophysical Research Letters}, 31(19):2--5, 2004.

\bibitem[ODD13]{OBrien2013}
L.~O'Brien, J.~M. Dudley, and F.~Dias.
\newblock {Extreme wave events in Ireland: 14 680 BP -- 2012}.
\newblock {\em Natural Hazards and Earth System Science}, 13(3):625--648, 2013.

\bibitem[Oka85]{Okada1985}
B.~Okada.
\newblock {Surface deformation due to shear and tensile faults in a
  half-space}.
\newblock {\em Bulletin of the Seismological Society of America}, 75(4):1135,
  1985.

\bibitem[RGG{\etalchar{+}}18]{Reguly2018}
I.Z. Reguly, D.~Giles, D.~Gopinathan, L.~Quivy, J.~H. Beck, M.~B. Giles,
  S.~Guillas, and F.~Dias.
\newblock {The VOLNA-OP2 tsunami code (version 1.5)}.
\newblock {\em Geoscientific Model Development}, 11(11):4621--4635, 2018.

\bibitem[SCC05]{Semmane2005}
F.~Semmane, M.~Campillo, and F.~Cotton.
\newblock {Fault location and source process of the Boumerdes, Algeria,
  earthquake inferred from geodetic and strong motion data}.
\newblock {\em Geophysical Research Letters}, 32(1):1--4, 2005.

\bibitem[SGGD17]{Salmanidou2017}
D.~M. Salmanidou, S.~Guillas, A.~Georgiopoulou, and F.~Dias.
\newblock {Statistical emulation of landslide-induced tsunamis at the Rockall
  Bank, NE Atlantic}.
\newblock {\em Proceedings of the Royal Society A: Mathematical, Physical and
  Engineering Sciences}, 473(2200), 2017.

\bibitem[WC94]{Wells1994}
D.~L. Wells and K.~J. Coppersmith.
\newblock {New empirical relationships among magnitude, rupture length, rupture
  width, rupture area, and surface displacement}.
\newblock {\em Bulletin - Seismological Society of America}, 84(4):974--1002,
  1994.

\bibitem[YLM{\etalchar{+}}04]{Yelles2004}
K.~Yelles, K.~Lammali, A.~Mahsas, E.~Calais, and P.~Briole.
\newblock {Coseismic deformation of the May 21st, 2003, Mw = 6.8 Boumerdes
  earthquake, Algeria, from GPS measurements}.
\newblock {\em Geophysical Research Letters}, 31(13), 2004.

\end{thebibliography}

\end{document}